\documentclass[superscriptaddress,12pt, preprint, a4paper,nofootinbib]{revtex4}
\usepackage{amsmath, amssymb, slashed, epsf, color}
\usepackage{multirow}
\usepackage{here}
\usepackage{graphicx}
\usepackage{cancel}
\usepackage{tikz}

\begin{document}
\renewcommand{\thefootnote}{\fnsymbol{footnote}}
\preprint{UT-HET-111}
\preprint{CTPU-16-05}
\preprint{KU-PH-020}

\title{Single and double production of the Higgs boson \\ 
at hadron and lepton colliders 
in minimal composite Higgs models}
	\author{Shinya~Kanemura}
	\email{kanemu@sci.u-toyama.ac.jp}
	\affiliation{Department of Physics, University of Toyama,\\ 
	3190 Gofuku, Toyama, 930-8555, Japan}
	\author{Kunio~Kaneta}
	\email{kaneta@ibs.re.kr}
	\affiliation{Center for Theoretical Physics of the Universe, Institute for Basic Science (IBS), Daejeon 34051, Republic of Korea
}
	\author{Naoki~Machida}
	\email{machida@jodo.sci.u-toyama.ac.jp}
	\affiliation{Department of Physics, University of Toyama,\\ 
	3190 Gofuku, Toyama, 930-8555, Japan}
	\author{Shinya~Odori}
	\email{odri@jodo.sci.u-toyama.ac.jp}
	\affiliation{Department of Physics, University of Toyama,\\ 
	3190 Gofuku, Toyama, 930-8555, Japan}
	\author{Tetsuo~Shindou}
	\email{shindou@cc.kogakuin.ac.jp}
	\affiliation{Division of Liberal-Arts and Department of Applied Physics, Kogakuin University, \\
	1-24-2 Nishi-Shinjuku, Tokyo, 163-8677, Japan}
	
	\begin{abstract}
		In the composite Higgs models, originally proposed by Georgi and Kaplan, the Higgs boson is a pseudo Nambu-Goldstone boson (pNGB)
		of spontaneous breaking of a global symmetry.
		In the minimal version of such models, global SO(5) symmetry is spontaneously broken to SO(4), and 
		the pNGBs form an isospin doublet field, which corresponds to the 
		Higgs doublet in the Standard Model (SM).
		Predicted coupling constants of the Higgs boson can in general deviate from the SM predictions, depending on the compositeness parameter.  
		The deviation pattern is determined also by the detail of 
		the matter sector.
		We comprehensively study how the model can be tested via measuring single and 
		double production processes of the Higgs boson
				at LHC and future electron-positron colliders.
		The possibility to distinguish the matter sector among the minimal composite
		Higgs models is also discussed.
		In addition, we point out differences in the cross section of
		double Higgs boson production from the prediction in other new physics models.
				 
		 \end{abstract}

	\setcounter{footnote}{0}
	\renewcommand{\thefootnote}{\arabic{footnote}}
	\maketitle

\section{Introduction}

By measurements of the Higgs boson property at the LHC Run-I\cite{HiggsatLHC,LHCRun1KVKF}, it has been found that 
the Standard Model (SM), including the sector of electroweak symmetry breaking, is surely
 a good description of the physics at the electroweak scale.
However the essence of the Higgs boson has not been understood yet, 
and it is worth asking whether the Higgs boson is really an elementary scalar or a composite state of more fundamental fields.

Although the idea of the Higgs boson as an elementary scalar field may be justified 
in the framework of supersymmetry (SUSY) at the TeV scale, 
predicted superpartner particles have not been discovered up to now at LHC.
In this situation, motivation for a scenario of the Higgs boson as a composite state
becomes stronger.

In particular, with the measured mass of the Higgs boson to be 125~GeV, 
it is interesting to consider the scenario originally proposed by Gerogi and Kaplan\cite{GeorgiKaplan}, 
where the Higgs boson is a pseudo Nambu-Goldstone boson (pNGB)
of spontaneous symmetry breaking of a global symmetry above the TeV scale.
Models based on this idea are called as composite Higgs models,
which draw a lot of attention recently\cite{MCHMs,MCHMreview}. 

The minimal composite Higgs model (MCHM)\cite{MCHMs} is known as 
a minimal realisation of the composite Higgs models, where the approximate global symmetry 
SO(5)$\times$ U(1)$_X$ is spontaneously broken into SO(4)$\times$U(1)$_X$ at the scale $f$. 
Associated with the symmetry breaking, 
four pNGBs appear, which are identified as the four component fields in 
the Higgs doublet of the SM.
There are several attempts to probe the MCHMs by phenomenological study 
at the present and future collider experiments\cite{MCHMpheno, Carena, KKMS,  MCHMPT, pphhMCHM, eehhMCHM}.

The induced Higgs potential depends on which representations of SO(5) the matter fermions are embedded into.
The Higgs boson couplings with the matter fermions such as Yukawa couplings are also 
determined by the representations of the matter fermions.
In the literature, different patterns of the deviations of the coupling constants 
from the SM predictions are shown in different MCHMs\cite{Carena, KKMS}.
Therefore, measuring the deviation pattern at collider experiments is critical to distinguish 
a variety of the MCHMs.

In this paper, we investigate collider phenomenology of the MCHMs.
Because of the deviations of the Higgs boson couplings from the SM prediction, 
the production cross section and the branching ratios of the Higgs boson 
are different from those in the SM.
We first clarify the constraint on the compositeness parameter in the MCHMs 
at the LHC Run-I. 
We then discuss double Higgs boson production at LHC and future electron-positron colliders,
taking into account the constraint. 

At the LHC Run-I, single Higgs boson production 
has been measured in various decay modes, and no significant deviation from the SM 
has been detected yet. 
In the literature\cite{ConstonXi}, 
the constraint on the compositeness parameter is obtained
in the case of the universal scale factors for the Yukawa couplings.
In models with non-universal scale factors, the constraint should be 
extracted from the signal strength of individual production and decay mode
measured at the LHC Run-I\cite{LHCRun1KVKF}.

With taking into account the constraint, we evaluate cross sections of double Higgs boson production at LHC with 
the collision energy of 14~TeV in several MCHMs, 
and discuss the possibility to distinguish among different MCHMs.
The search for double Higgs boson production process at LHC
can access to the coupling deviation which does not contribute to   
the single Higgs boson production.
While there have already been several phenomenological studies of the double Higgs boson 
production at LHC in the context of MCHMs\cite{pphhMCHM}, 
we here study the process from the viewpoint of the discrimination 
of the matter sector in various MCHMs.

The double Higgs boson production process at future electron-positron colliders such as 
the International Linear Collider (ILC)\cite{ILC} and the Compact LInear Collider (CLIC)\cite{CLIC}
can give complementary information of the new physics. 
At the electron-positron colliders, the double Higgs boson production cross section 
can be measured as a function of the collision energy.
It is known that the production cross section has a specific 
energy dependence on new physics models\cite{Asakawa:2010xj}.
We estimate production cross sections for double Higgs boson 
production processes in MCHMs, 
and show their energy dependences.

This article is organised as follows. 
In Sec.~II, we give a brief review of the framework of the MCHM. We introduce three 
different MCHMs as benchmark models, and list deviation patterns in Higgs boson coupling constants from the SM predictions and decay branching ratios of the Higgs boson.
In Sec.~III, we show the present bound on the model parameter from 
the LHC Run-I data, and we discuss the double Higgs boson production process at LHC
and a future electron-positron collider experiment.
Summary and conclusion are shown in Sec.~IV.

\section{Higgs boson couplings in various MCHMs}
\label{sec:review}

In this section, we give a brief review on the MCHM for self consistency.
In the MCHM\cite{MCHMs,MCHMreview}, four components of the SM-like Higgs boson doublet are identified 
with the four pNGBs associated with the spontaneous symmetry breaking of 
$\text{SO}(5)\times \text{U}(1)_X\to \text{SO}(4)\times \text{U}(1)_X$.
The low energy effective Lagrangian can be 
constructed by utilising a nonlinear representation of the global symmetry 
group.
Since $\text{SU}(2)_L\times \text{U}(1)_Y$ in $\text{SO}(4)\times \text{U}(1)_X$ 
is gauged, the global symmetry is explicitly broken by the 
gauge couplings.
The matter fermions in the SM by themselves cannot consist 
$\text{SO}(5)$ multiplet. They should be embedded
into $\text{SO}(5)$ representations,  
and the components of the multiplets other than the SM fermions are 
decoupled from the effective theory. The Yukawa coupling with the matter
fermions of the SM also explicitly breaks the global $\text{SO}(5)$
symmetry.
Due to these explicit breaking effects of the global symmetry, 
the Higgs potential is effectively generated via the loop 
contributions of weak gauge bosons and 
the SM fermions. 
The SM fermion contribution $V_{\text{eff}}^{\text{fermion}}$ strongly depends on 
how the SM fermion fields are embedded into 
$\text{SO}(5)$ multiplets.
The detail of the model dependence of the Higgs 
potential in the MCHMs are given in Ref.~\cite{Carena,KKMS}.
By solving the stationary condition, 
the Higgs field can have non-zero vacuum expectation value (VEV).

We summarise deviations in the  Higgs boson couplings from the SM predictions
in the different MCHMs.
In order to describe the deviation pattern in each MCHM, it is convenient to use 
the scale factors $\kappa_a$, which are defined by $\kappa_a\equiv g_a/g_a^{\text{SM}}$, where $g_a$ denote the
coupling constants of the Higgs boson coupling with the weak gauge bosons $V=W$ and $Z$, matter fermions
and the Higgs boson itself as $a=hVV$, $htt$, $hbb$, and $hhh$.
For $\kappa_{hVV}$, $\kappa_{htt}$ and 
$\kappa_{hbb}$, we simply write $\kappa_V$, $\kappa_t$ and $\kappa_b$, respectively.
For $hhVV$ couplings, we introduce the parameter 
$c_{hhVV}=g_{hhVV}/g_{hhVV}^{\text{SM}}$.
In the effective theories of the MCHMs, new dimension five operators of two Higgs bosons and two fermions 
such as $hht\bar{t}$ are induced. We parameterise the coupling constant 
for $hht\bar{t}$ as $g_{hhtt}=c_{hhtt}m_t/(2v^2)$.
The four point interactions such as $hhVV$ and $hht\bar{t}$ play important roles in
our analysis.

In the MCHMs, the couplings between the Higgs boson and the weak 
gauge bosons deviate from the SM predictions as $g_{hVV}=g_{hVV}^{\text{SM}}\sqrt{1-\xi}$, 
which leads to $\kappa_V=\sqrt{1-\xi}$,
no matter how the SM fermions are embedded into $\text{SO}(5)$ representations.
It is the universal prediction of the MCHMs.
Here the compositeness parameter $\xi$ is defined as $\xi\equiv v^2/f^2$ 
where $v(\simeq 246~\text{GeV})$ is the VEV of the electroweak symmetry breaking.
The coupling constants of the contact interaction $hhVV$ are also determined universal in the
MCHMs as $c_{hhVV}=1-2\xi$.
The universal reduction of $c_{hhVV}$ is a notable feature of the MCHMs.
In the Higgs sector with multi-doublet structure such as that in the minimal SUSY SM, 
the coupling constants $c_{hhVV}$ are given by the gauge coupling constants
as the same as those in the SM.
 
On the other hand, Yukawa coupling constants with the matter fermions as well as 
the self-coupling of the Higgs boson
are determined by the matter sector of the MCHMs.
As described above, the matter fermions of the SM are embedded into a representation of 
$\text{SO}(5)$.
We can individually embed $q_L=(u_L,d_L)$, $u_R$, $d_R$, $\ell_L=(\nu_L,e_L)$ and $e_R$.
Depending on the choices for charge assignment for fermions, a variety of the models can be
constructed.
The patterns of scale factors in various MCHMs are listed in Refs.~\cite{Carena,KKMS}.

In order to demonstrate how to distinguish the MCHMs, 
we pick up three typical models, MCHM$4$, MCHM$5$ and MCHM${14}$, 
in which all the matter fermions are embedded into the four-, five-, and fourteen-dimensional
representations of $\text{SO}(5)$, respectively.
The relevant matter part of the effective Lagrangian in each model is given by
\begin{align}
\mathcal{L}_{\text{MCHM}4}^{\text{matter}}=&
	\sum_{r=q,\ell,u,d,e}\overline{\Psi}_r^{(4)}\slashed{p}\left[\Pi_0^r+\Pi_1^r\Gamma^i\Sigma_i\right]\Psi_r^{(4)}
	+\overline{\Psi}_q^{(4)}\left[M_0^t+M_1^t\Gamma^i\Sigma_i\right]\Psi_t^{(4)}\nonumber\\
	&+\overline{\Psi}_q^{(4)}\left[M_0^b+M_1^b\Gamma^i\Sigma_i\right]\Psi_b^{(4)}
	+\overline{\Psi}_{\ell}^{(4)}\left[M_0^{\tau}+M_1^{\tau}\Gamma^i\Sigma_i\right]\Psi_{\tau}^{(4)}
	+\text{h.c.}
	\;,
	\nonumber\\
\mathcal{L}_{\text{MCHM}5}^{\text{matter}}=&
	\sum_{r=q,\ell,u,d,e}\overline{\Psi}_r^{(5)}\left[\slashed{p}\Pi_0^r+\Sigma^{\dagger}\slashed{p}\Pi_1^r\Sigma\right]\Psi_r^{(5)}
	+\overline{\Psi}_{q}^{(5)}\left[M_0^t+\Sigma^{\dagger}M_1^t\Sigma\right]\Psi_{t}^{(5)}\nonumber\\
	&
	+\overline{\Psi}_{q}^{(5)}\left[M_0^b+\Sigma^{\dagger}M_1^b\Sigma\right]\Psi_{b}^{(5)}
	+\overline{\Psi}_{\ell}^{(5)}\left[M_0^{\tau}+\Sigma^{\dagger}M_1^{\tau}\Sigma\right]\Psi_{{\tau}}^{(5)}
	+
	\text{h.c.}\;,
	\nonumber\\
	\mathcal{L}_{\text{MCHM}{14}}^{\text{matter}}=&
	\sum_{r=q,\ell,u,d,e}\left[\overline{\Psi}_r^{(14)}\slashed{p}\Pi_0^r\Psi_r^{(14)}
		+(\Sigma\overline{\Psi}_r^{(14)})\slashed{p}\Pi_1^r(\Psi_r^{(14)}\Sigma^{\dagger})
	+(\Sigma\overline{\Psi}_r^{(14)}\Sigma^{\dagger})\slashed{p}\Pi_2^r(\Sigma\Psi_r^{(14)}\Sigma^{\dagger})\right]
	\nonumber\\
	&
	+\overline{\Psi}_{q}^{(14)}M_0^t\Psi_{t}^{(14)}
	+(\Sigma\overline{\Psi}_{q}^{(14)})M_1^t(\Psi_{t}^{(14)}\Sigma^{\dagger})
	+(\Sigma\overline{\Psi}_{q}^{(14)}\Sigma^{\dagger})M_2^t(\Sigma\Psi_{t}^{(14)}\Sigma^{\dagger})\nonumber\\
	&
	+\overline{\Psi}_{q}^{(14)}M_0^b\Psi_{b}^{(14)}
	+(\Sigma\overline{\Psi}_{q}^{(14)})M_1^b(\Psi_{b}^{(14)}\Sigma^{\dagger})
	+(\Sigma\overline{\Psi}_{q}^{(14)}\Sigma^{\dagger})M_2^b(\Sigma\Psi_{b}^{(14)}\Sigma^{\dagger})\nonumber\\
	&+\overline{\Psi}_{\ell}^{(14)}M_0^{\tau}\Psi_{\tau_R}^{(14)}
	+(\Sigma\overline{\Psi}_{\ell}^{(14)})M_1^{\tau}(\Psi_{\tau}^{(14)}\Sigma^{\dagger})
	+(\Sigma\overline{\Psi}_{\ell}^{(14)}\Sigma^{\dagger})M_2^{\tau}(\Sigma\Psi_{\tau}^{(14)}\Sigma^{\dagger})
	+\text{h.c.}\;,
	\label{eq:matterLag}
\end{align}
where $\Sigma$ is given by 
\begin{equation}
\Sigma=\frac{\sin(h/f)}{h}(h^1,h^2,h^3,h^4,h\cot(h/f))\;,\quad h=\sqrt{h^ah^a}\;,	
\end{equation}
with $h^a$ being the pNGBs\footnote{We take $h^3$ as the physical Higgs boson\cite{MCHMs,MCHMreview}.}, 
$\Psi_r^{(R)}$ denotes the $R$-dimensional representation into which the SM matter fermion $r$ is embedded, 
$\Gamma_i$ are the gamma matrices in SO(5), 
and $\Pi$'s and $M$'s are the form factor.
In the analysis, we focus on the third generation of the fermions.
With this Lagrangian, 
the Higgs potential and the Yukawa couplings of the fermions can be computed in each model.

In Table~\ref{Table:fp}, we summarise a set of scale factors in
the three models up to the linear term of $\xi$.
Since the compositeness parameter $\xi$ should 
be much smaller than unity by the experimental constraints as we discuss later, 
the higher order terms are negligible.  
In MCHM${14}$, some scale factors depend not only on $\xi$ but also on $M_1^t/M_2^t$.
In the limit of $M_2^t\to 0$, the deviation pattern of the scale factors becomes the same as that in 
MCHM$5$.
In the following analysis, we set $M_1^t=0$ for MCHM${14}$ for simplicity.
\begin{table}[t]
\caption{
Deviations in coupling constants with 
the Higgs boson in MCHM$4$, MCHM$5$ and MCHM${14}$.
We here show the formulae for $\xi\ll 1$.
The scale factors for more variety of MCHMs are given in Refs.~\cite{Carena,KKMS}.
\label{Table:fp}
}	
\begin{tabular}{|l|c|c|c|c|c|c|c|c|c|c|}
 \hline
Model& $\kappa_{V}$  & $c_{hhVV}$ & $\kappa_{hhh}$ & $\kappa_{t}$ & $\kappa_{b}$&$\kappa_{\tau}$& $c_{hhtt}$ \\ \hline \hline
${\rm MCHM}{4}$                                                                                                                                                                                                                         
                 & \multirow{3}{*}{$1-\frac{1}{2}\xi$} & \multirow{3}{*}{$1-2\xi$} & $1-\frac{1}{2}\xi$ & $1-\frac{1}{2}\xi$ & $1-\frac{1}{2}\xi$ &$1-\frac{1}{2}\xi$& $-\xi$  \\ \cline{1-1}\cline{4-8} 
${\rm MCHM}{5}$ & &  & \multirow{2}{*}{$1-\frac{3}{2}\xi$} & 
$1-\frac{3}{2}\xi$ & \multirow{2}{*}{$1-\frac{3}{2}\xi$} &\multirow{2}{*}{$1-\frac{3}{2}\xi$}&$-4\xi$ \\ 
\cline{1-1}\cline{5-5}\cline{8-8}
${\rm MCHM}{14}$ &   & &  &                                                                                                                             
                           $1-\frac{9M_1^t+64M_2^t}{6M_1^t+16M_2^t}\xi$ & & & $-\frac{4(3M_1^t+23M_2^t)}{3M_1^t+8M_2^t}\xi$ \\ 
                                                   \hline 
\end{tabular}
\end{table}

\begin{figure}[t]
\begin{tabular}{cc}
	\includegraphics[scale=0.8]{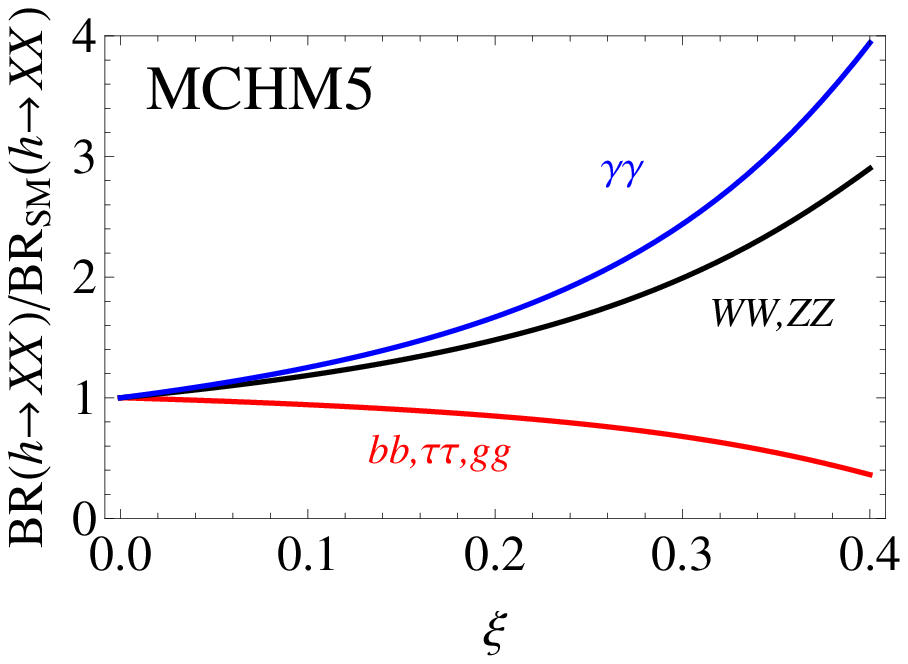}&
	\includegraphics[scale=0.8]{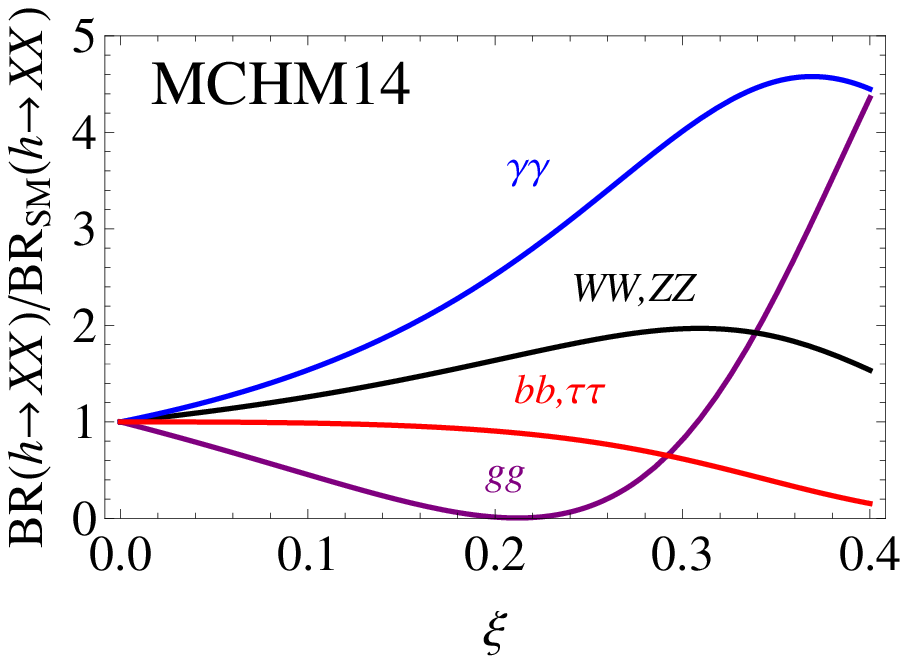}\\
\end{tabular}
\caption{
Ratios of the Higgs boson decay branching ratios	
in MCHM$5$, and MCHM${14}$
to those in the SM
as a function of the compositeness parameter $\xi$.
\label{fig:Br}}
\end{figure}

The deviations in the Higgs boson couplings from the SM predictions 
affect the Higgs boson decay branching ratios.
In Fig.~\ref{fig:Br}, we show ratios of decay branching ratios of the Higgs boson in 
MCHM$5$ and MCHM${14}$ to the SM prediction in each decay mode.
In the figure, $\text{BR}(h\to XX)$  
is the decay branching ratio of $h\to XX$ mode 
 ($XX=ZZ,WW,\gamma\gamma,\bar{\tau}\tau$ and $\bar{b}b$)
and $\text{BR}_{\text{SM}}(h\to XX)$ is its SM prediction. 
In Table~\ref{BrinMCHMs}, the decay branching ratios in the SM and in the MCHMs with typical value of $\xi$ are shown.
In MCHM$4$, all the decay widths are shifted by the same factor $\kappa_V^2=\kappa_f^2=1-\xi$. 
The decay branching ratios are then the same as those in the SM.
In numerical evaluation, we take into account the NLO QCD corrections\cite{NLOQCDinBr} and the NLO electroweak corrections\cite{NLOEWinBr} to the branching ratios.
\begin{table}[t]
\caption{The decay branching ratios of the Higgs boson in the SM,  MCHM$4$, MCHM$5$, and MCHM${14}$. 
For MCHM$5$ and MCHM${14}$, the compositeness parameter $\xi$ is taken as  
$\xi=0.1$ and $0.2$.
The NLO QCD corrections\cite{NLOQCDinBr} and the NLO electroweak corrections\cite{NLOEWinBr} are taken into account.\label{BrinMCHMs}
}
\renewcommand{\arraystretch}{1.2}
\begin{tabular}{|c|c|c|c|c|c|c|} 
\hline
Branching Ratio          & $h \to bb$ & $h \to W W$ & $h\to gg$ & $h \to \tau \tau$  & $h \to ZZ$ & $h \to \gamma \gamma$ \\ \hline
SM and MCHM$4$           & 0.555     & 0.231     & 0.0894       & 0.0674            & 0.0282     &  0.00244          \\ \hline \hline
MCHM$5$ ($\xi = 0.1$)   & 0.517     & 0.274      &0.0843      & 0.0629            & 0.0334     & 0.00305              \\ 
MCHM$5$ ($\xi = 0.2$)   & 0.457     & 0.343      &0.0759      & 0.0556            & 0.0417     & 0.00407             \\\hline
MCHM${14}$ ($\xi = 0.1$)& 0.548     & 0.292      &0.0406      & 0.0668            & 0.0355     & 0.00374             \\ 
MCHM${14}$ ($\xi = 0.2$)& 0.502     & 0.379      &0.00121      & 0.0615            & 0.0461     & 0.00615             \\ \hline
\end{tabular}
\end{table}

\section{Phenomenology of the MCHMs at collider experiments}
\subsection{Current constraint on the compositeness parameter from the LHC Run-I data}
It is known that the value of the compositeness parameter $\xi$ 
is constrained as $\xi <0.25$ from the electroweak precision data\cite{MCHMreview,MCHMPT}.
We here consider the constraint from the LHC Run-I data, which gives a significant impact.
In order to take into account the constraint from the data without extra assumption, 
we compare the model prediction with the measurement 
by using the signal strength defined as 
\begin{equation}
\mu_I^{XX}=\frac{\sigma_I\cdot \text{BR}(h\to XX)}{(\sigma_I)_{\text{SM}}\cdot \text{BR}_{\text{SM}}(h\to XX)}\;,
\end{equation}
where $\sigma_I$  is the production cross section in several processes as $I=\text{ggF(gluon fusion)}$, $\text{VBF(vector boson fusion)}$, $Wh$, $Zh$ and $tth$(associate productions with $W$, $Z$, and $\bar{t}t$ respectively), and $(\sigma_I)_{\text{SM}}$ is its SM prediction.
The measured values of the signal strength in various modes with error are listed in 
Table~\ref{signalstrength}, which are taken from Ref.~\cite{LHCRun1KVKF}.
In this table, $\mu_V^{XX}$ and $\mu_F^{XX}$ are the abbreviations for $\mu_{\text{VBF}+Vh}^{XX}$ and 
$\mu_{\text{ggF}+tth}^{XX}$, respectively. 
In the MCHMs, the prediction on the signal strength $\mu_V^{XX}$ and $\mu_{F}^{XX}$ are given by 
\begin{align}
\mu_{V}^{XX}=\kappa_V^2\frac{\text{BR}(h\to XX)}{\text{BR}_{\text{SM}}(h\to XX)}\;,\quad
\mu_{F}^{XX}=\kappa_t^2\frac{\text{BR}(h\to XX)}{\text{BR}_{\text{SM}}(h\to XX)}\;.	
\end{align}

\begin{table}[t]
\caption{The signal strength in various modes measured at ATLAS and CMS. The 
values are taken from the results for the 10-parameter fit in Ref.~\cite{LHCRun1KVKF}.
\label{signalstrength}
}
\begin{tabular}{c|c|c|c}\hline
&ATLAS+CMS&ATLAS&CMS\\\hline
$\mu_{F}^{\gamma\gamma}$&$1.19^{+0.28}_{-0.25}$&$1.31^{+0.37}_{-0.34}$&$1.01^{+0.34}_{-0.31}$\\
$\mu_{F}^{ZZ}$&$1.44^{+0.38}_{-0.34}$&$1.73^{+0.51}_{-0.45}$&$0.97^{+0.54}_{-0.42}$\\
$\mu_{F}^{WW}$&$1.00^{+0.23}_{-0.20}$&$1.10^{+0.29}_{-0.26}$&$0.85^{+0.28}_{-0.25}$\\
$\mu_{F}^{\bar{\tau}\tau}$&$1.10^{+0.61}_{-0.58}$&$1.72^{+1.24}_{-1.13}$&$0.91^{+0.69}_{-0.64}$\\
$\mu_{F}^{\bar{b}b}$&$1.09^{+0.93}_{-0.89}$&$1.51^{+1.15}_{-1.08}$&$0.10^{+1.83}_{-1.86}$\\ 
$\mu_{V}^{\gamma\gamma}$&$1.05^{+0.44}_{-0.41}$&$0.69^{+0.64}_{-0.58}$&$1.37^{+0.62}_{-0.56}$\\
$\mu_{V}^{ZZ}$&$0.48^{+1.37}_{-0.91}$&$0.26^{+1.60}_{-0.91}$&$1.44^{+2.32}_{-2.30}$\\
$\mu_{V}^{WW}$&$1.38^{+0.41}_{-0.37}$&$1.56^{+0.52}_{-0.46}$&$1.08^{+0.65}_{-0.58}$\\
$\mu_{V}^{\bar{\tau}\tau}$&$1.12^{+0.37}_{-0.35}$&$1.29^{+0.58}_{-0.53}$&$0.87^{+0.49}_{-0.45}$\\
$\mu_{V}^{\bar{b}b}$&$0.65^{+0.30}_{-0.29}$&$0.50^{+0.39}_{-0.37}$&$0.85^{+0.47}_{-0.44}$\\\hline
\end{tabular}	
\end{table}

\begin{figure}[t]
\begin{tabular}{cc}
\includegraphics[scale=0.8]{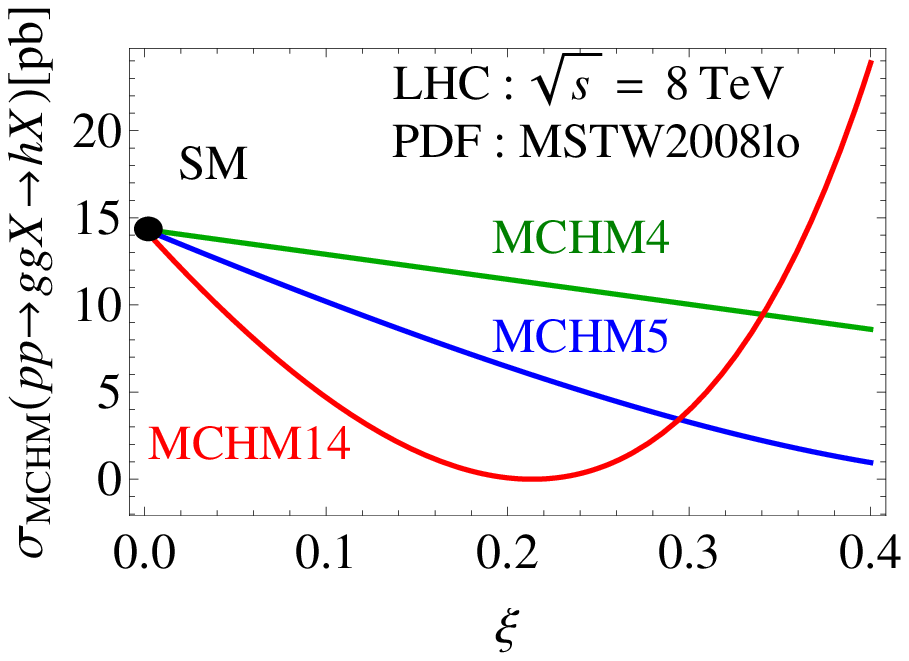}
&
\includegraphics[scale=0.8]{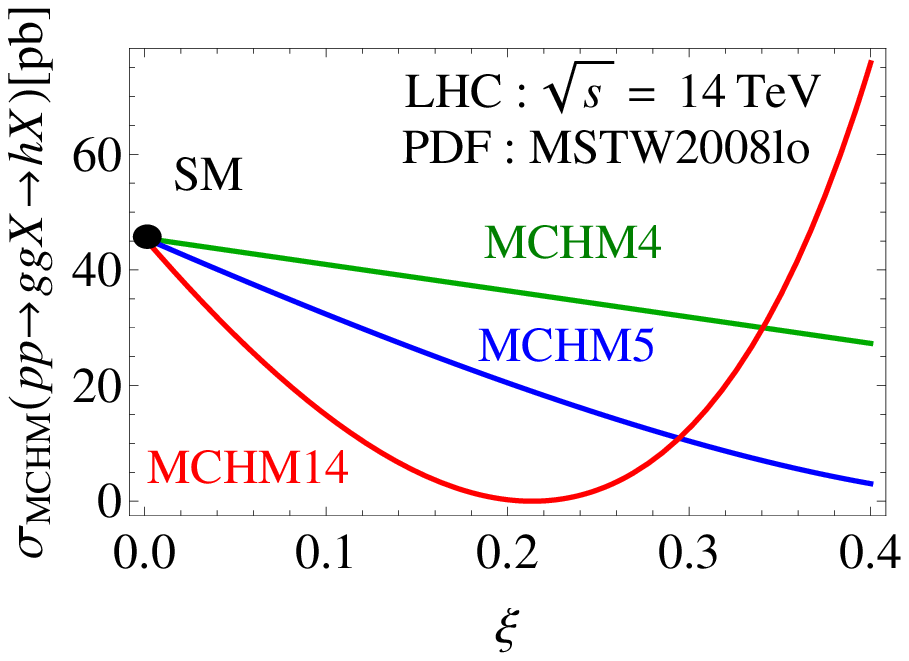}\\
(a)&(b)\\
\includegraphics[scale=0.8]{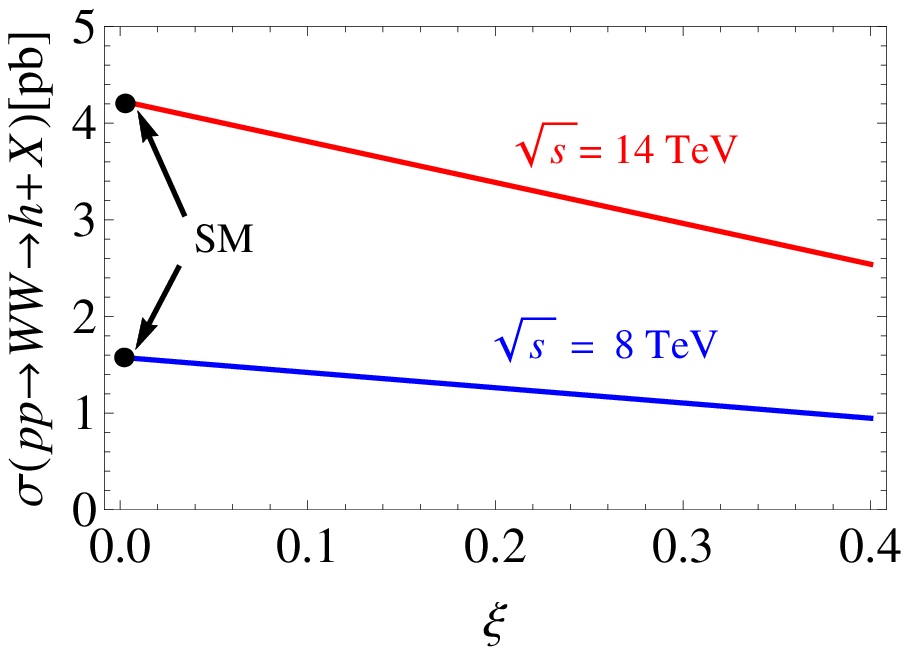}&\\
(c)&\\
\end{tabular}
\caption{(a) and (b): The production cross section of the single Higgs boson production via 
gluon fusion $pp\to ggX\to hX$ at LHC with (a) 8~TeV and (b) 14~TeV 
in MCHM$4$(Green), MCHM$5$(Blue) and MCHM${14}$(Red) 
as a function of $\xi$.
We use the $k$-factor as $k =2.38$ for the 8~TeV case and $k =2.27$ for the 14~TeV case.
(c): the production cross section of the single Higgs boson via $W$-fusion 
$pp\to W^+W^-X\to hX$ at LHC with 8~TeV (blue) and 14~TeV (red).
The cross section via $W$-fusion is universal for the MCHMs.
\label{Fig:SHP_XS}}
\end{figure}

In Fig.~\ref{Fig:SHP_XS}, we show the cross section of the single Higgs boson production $pp\to hX$ via gluon fusion 
and vector boson fusion at the LHC with 8~TeV and 14~TeV.
In the calculation for gluon fusion, we use the parton distribution function (PDF) of 
MSTW 2008lo\cite{PDF} and 
the NNLO $k$-factor as $k=2.38$ for 8~TeV 
and $k=2.27$ for 14~TeV\cite{deFlorian:2013jea}.
The production cross section for vector boson fusion with NNLO QCD correction and the NLO electroweak correction is predicted in the SM as 
$(\sigma_{\text{VBF}})_{\text{SM}} = 1.58~\text{pb}$ for the 8~TeV case 
and 
$(\sigma_{\text{VBF}})_{\text{SM}} = 4.23~\text{pb}$ for the 14~TeV case\cite{Higgs_XS}.
The gluon fusion process is determined by $\kappa_t$ so that the cross section depends on the models.
On the other hand the vector boson fusion process is determined by the value of 
$\kappa_V$. The cross section for the vector boson fusion is independent of the matter sector of the MCHMs.

\begin{figure}[t]
\begin{tabular}{cc}
\includegraphics[width=80mm]{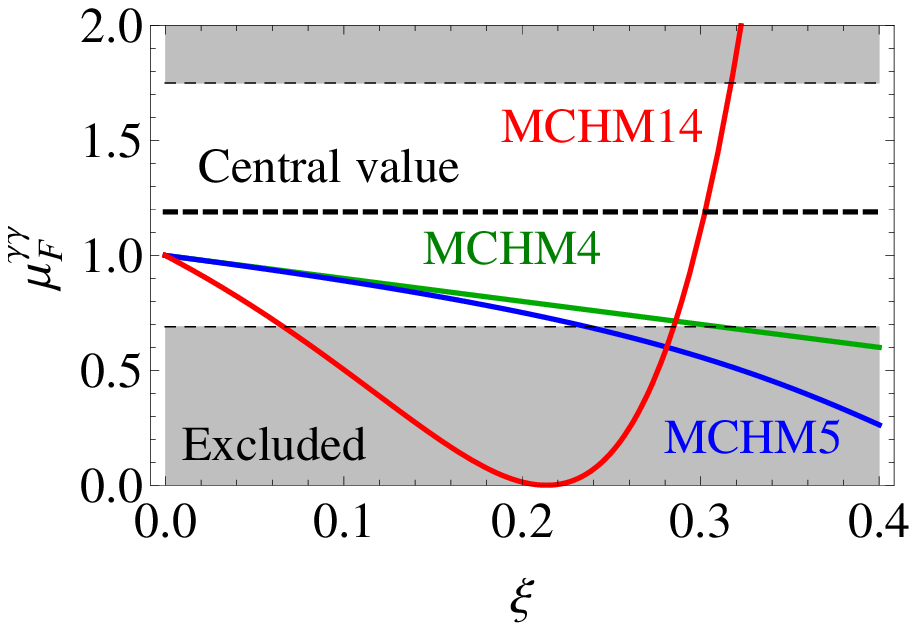}&
\includegraphics[width=80mm]{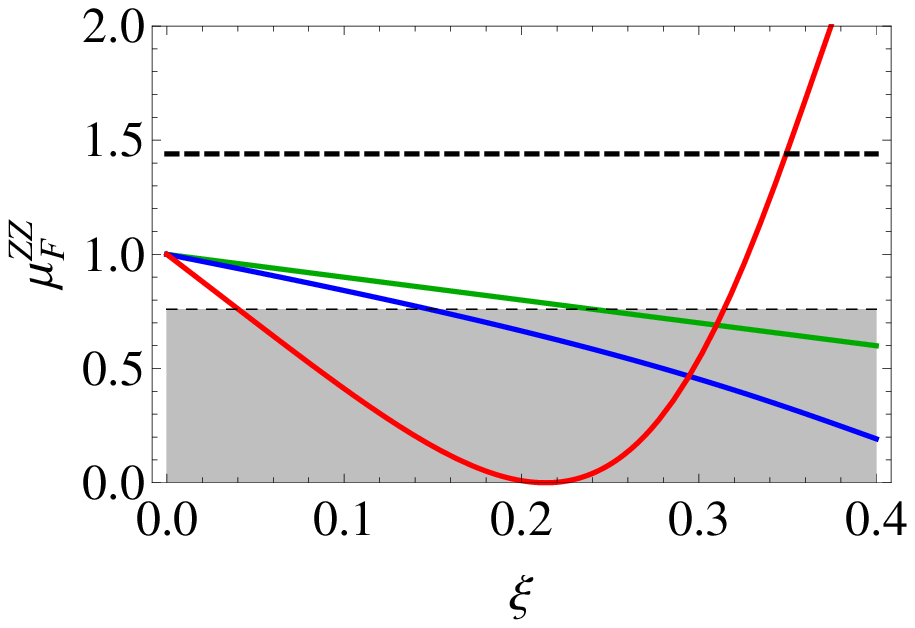}\\
\includegraphics[width=80mm]{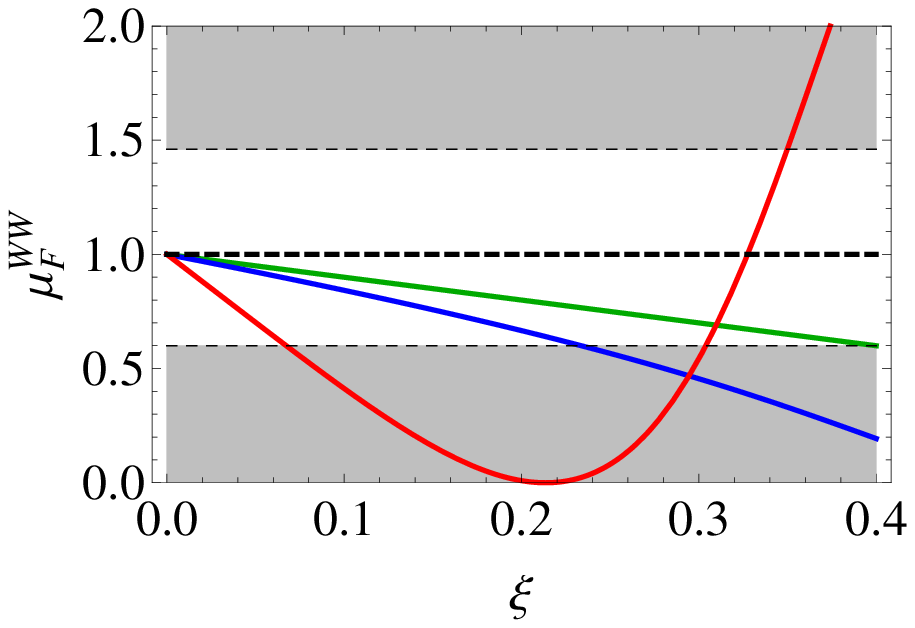}&
\includegraphics[width=80mm]{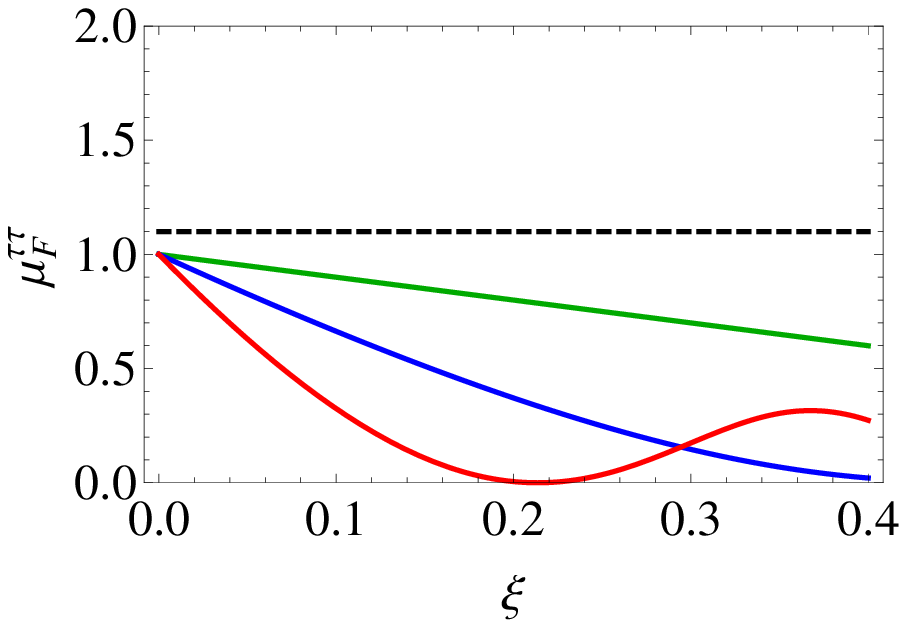}\\
\includegraphics[width=80mm]{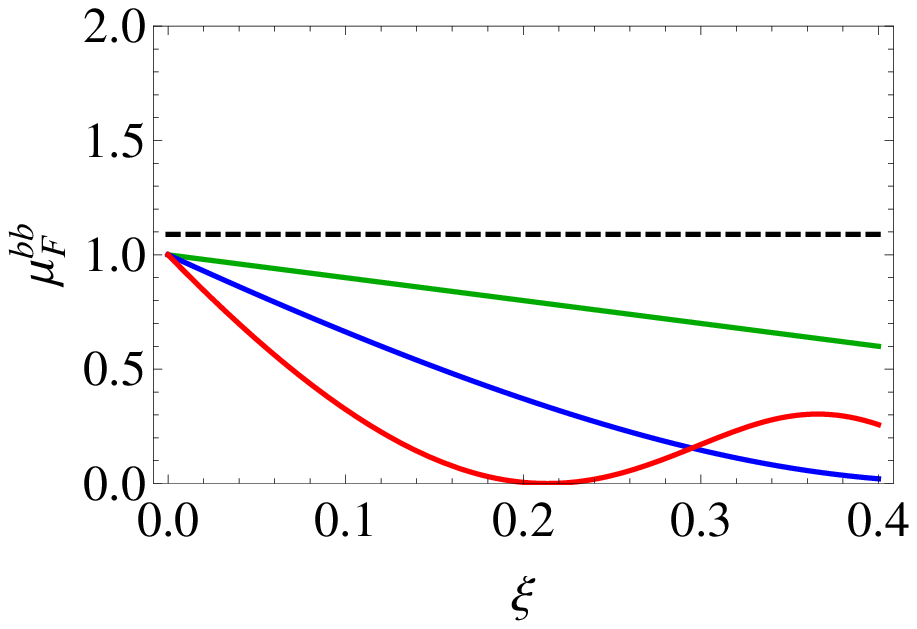}&\\
\end{tabular}
\caption{
The signal strength for the ggF+$tth$ production and various decay modes $\mu_F^{f}$ are shown.
The shaded area represents 2$\sigma$ excluded region given by the LHC Run-I data\cite{LHCRun1KVKF} in each figure.
\label{Fig:muF}}
\end{figure}

\begin{figure}[t]
\begin{tabular}{cc}
\includegraphics[width=80mm]{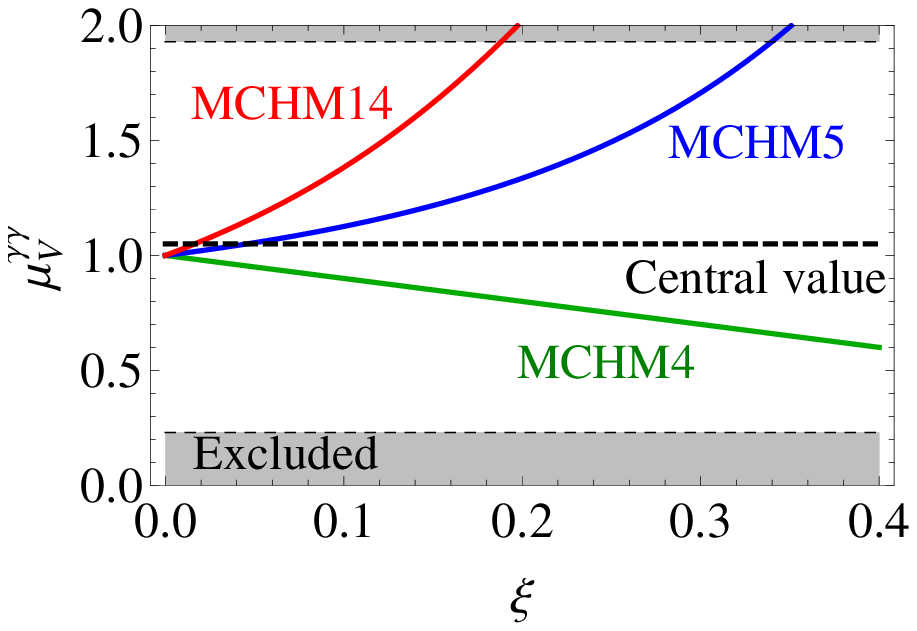}&
\includegraphics[width=80mm]{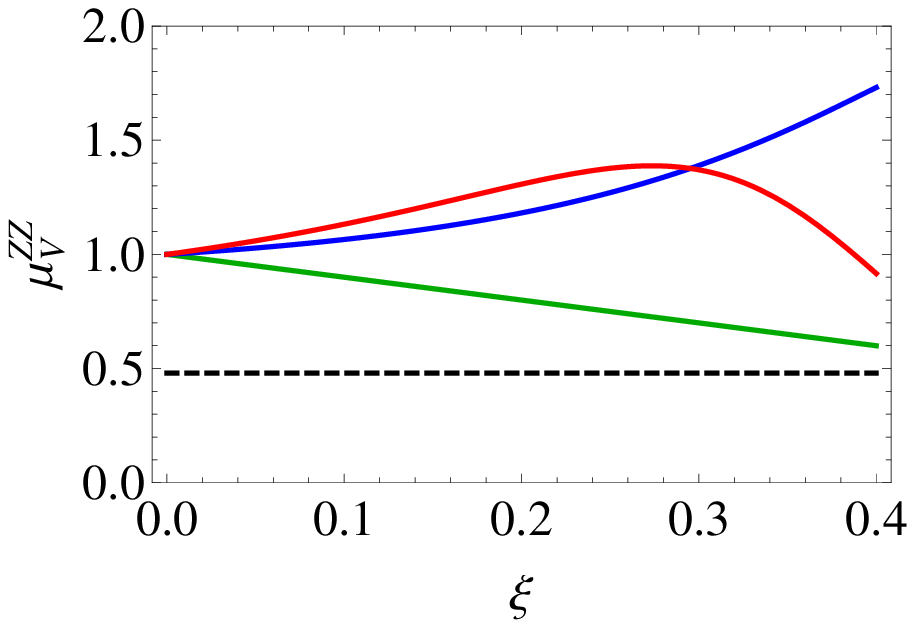}\\
\includegraphics[width=80mm]{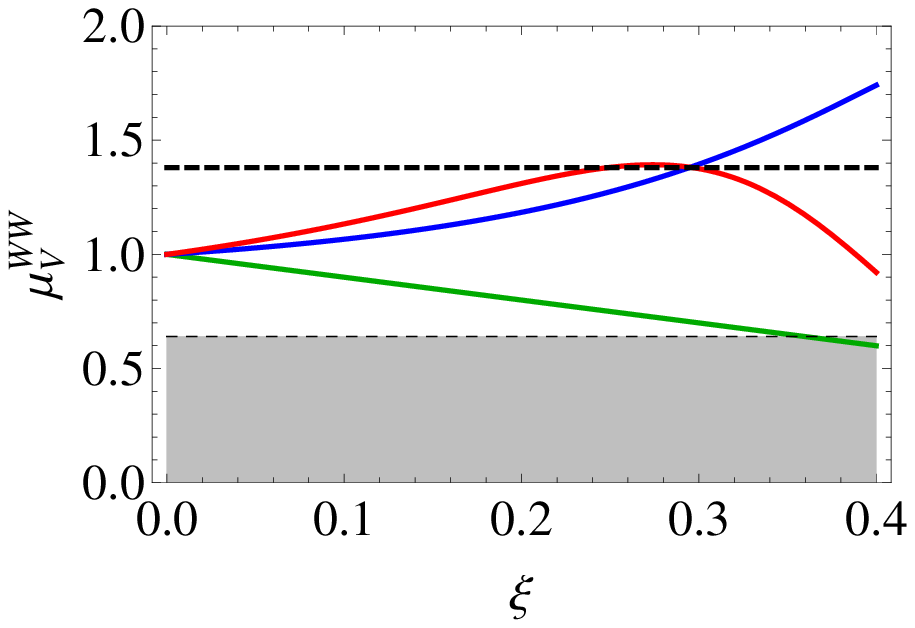}&
\includegraphics[width=80mm]{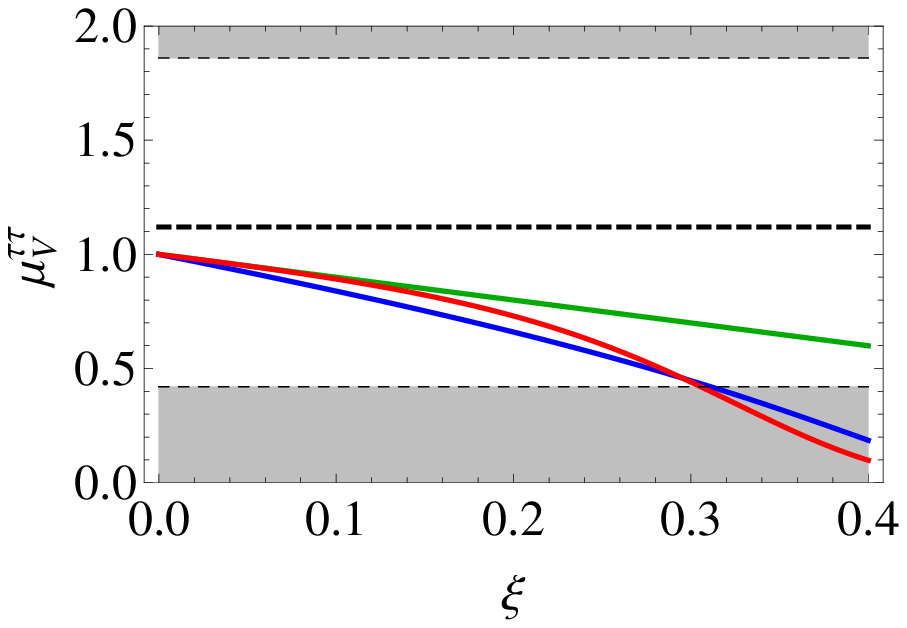}\\
\includegraphics[width=80mm]{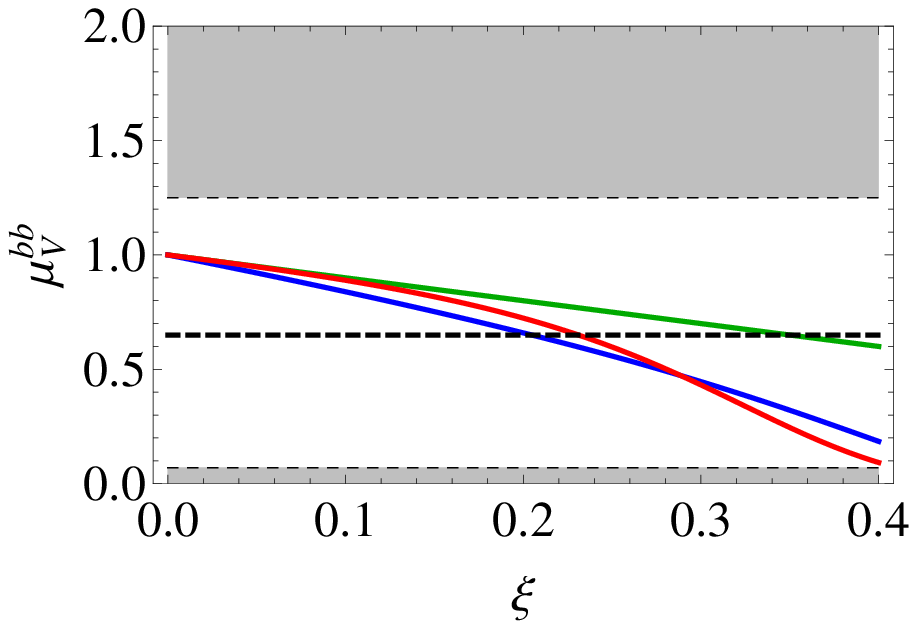}&\\
\end{tabular}
\caption{
The signal strength for the VBF+$Vh$ production and various decay modes $\mu_V^{f}$ are shown.
The shaded area represents 2$\sigma$ excluded region
 given by the LHC Run-I data\cite{LHCRun1KVKF}
 in each figure.
\label{Fig:muV}}
\end{figure}

In Figs.~\ref{Fig:muF} and \ref{Fig:muV}, 
values of the signal strength predicted in MCHM$4$, MCHM$5$ and MCHM${14}$ are shown. 
Comparing the predictions for the signal strength with the combined data of ATLAS and CMS\cite{LHCRun1KVKF}, 
we find significant constraints on the $\xi$ for each model from $\mu_F^{WW}$, $\mu_F^{ZZ}$, $\mu_F^{\gamma\gamma}$, 
$\mu_V^{WW}$, $\mu_V^{\gamma\gamma}$ and $\mu_V^{\tau\tau}$
as shown in  Table~\ref{constraintxi}.
MCHM${14}$ is strongly constrained, because the scale factor $\kappa_t$ decreases 
rapidly when $\xi$ becomes large.
In the three MCHMs, the values of the signal strength $\mu_F^{XX}$ are 
smaller than unity, 
while the measured values tend to be larger. 
In particular, the central value of $\mu_{F}^{ZZ}$ has slightly strong 
tension to the MCHMs prediction, so that this mode gives the strongest constraint
on the compositeness parameter $\xi$, in spite of the relatively large experimental uncertainty.
In the following discussion, we consider the parameter region of $\xi<0.25$ for MCHM$4$ and MCHM$5$ and 
$\xi<0.1$ for MCHM${14}$.

At the LHC Run-II, expected accuracy of each signal strength is more improved
to be within 10~\% for main channels such as 
$h\to \gamma\gamma, ZZ, WW, \tau\tau, bb$\cite{LHCRun2Precision}, 
so that the value of $\xi$ becomes more significantly constrained.


\begin{table}[t]
\caption{The 2$\sigma$ constraint on the value of $\xi$ 
obtained from the ATLAS and CMS combined data at the LHC Run-I experiments in MCHM$4$, MCHM$5$, and MCHM${14}$.
\label{constraintxi}}
\begin{tabular}{|c|c|c|c|c|c|c|}\hline
Model&$\mu_F^{\gamma\gamma}$&$\mu_F^{ZZ}$&$\mu_F^{WW}$&$\mu_V^{\gamma\gamma}$&$\mu_V^{WW}$&$\mu_V^{\tau\tau}$
\\ \hline
MCHM$4$&$\xi<0.31$&$\xi<0.24$&$\xi<0.40$&-&$\xi<0.36$&-
\\ \hline
MCHM$5$&$\xi<0.24$&$\xi<0.15$&$\xi<0.23$&$\xi<0.34$&-&$\xi<0.30$
\\ \hline 
MCHM${14}$&$\xi<0.07$&$\xi<0.04$&$\xi<0.07$&$\xi<0.19$&-&$\xi<0.30$
\\ \hline
\end{tabular}
\end{table}

\subsection{Double Higgs boson production at LHC}

Measuring the double Higgs boson production process provides an insight on the self coupling 
of the Higgs boson\cite{pphh}. At the High-Luminosity LHC (HL-LHC) experiment, 
the cross section of the double Higgs boson production is expected to be measured with 
54\% uncertainty in the SM case\cite{hhHLLHC}.
In new physics models,
the double Higgs boson production process at LHC is also important to explore the 
deviation pattern in the Higgs boson couplings.
Differently from the single Higgs boson production or the Higgs boson decay, 
contact interactions such as $c_{hhVV}$ and $c_{hhtt}$ 
play important roles.
The double Higgs boson production at LHC is analysed in the context of the MCHMs 
in the literature\cite{pphhMCHM}.
We here show our numerical results 
not only on the production cross section of 
the double Higgs boson production process, $pp\to hhX$, 
but also on the signal strength for each decay mode of $hh$
in MCHM$4$, MCHM$5$ and MCHM$14$.

\begin{figure}[t]
\begin{center}
\begin{tabular}{cccc}
\includegraphics[scale=0.7]{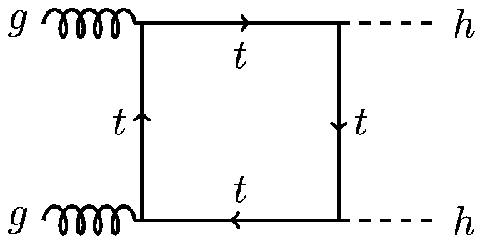}\phantom{Spa}&
\includegraphics[scale=0.7]{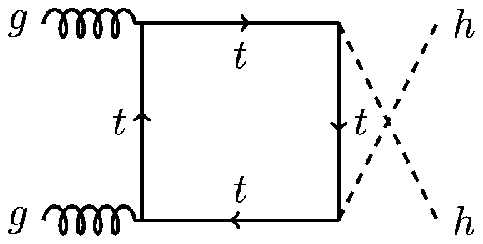}\phantom{Spa}&
\includegraphics[scale=0.7]{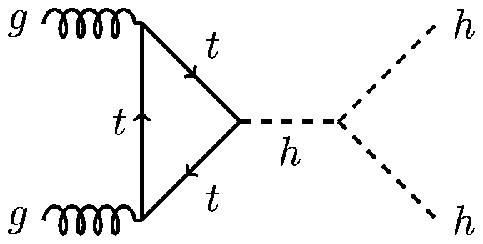}\phantom{Spa}&
\includegraphics[scale=0.7]{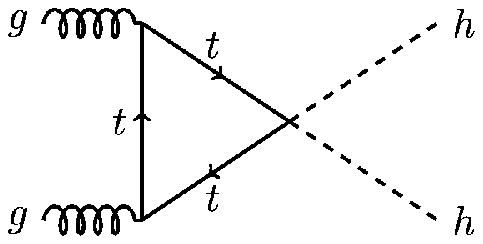}\\
(a)&(b)&(c)&(d)
\end{tabular}
\end{center}
\caption{Feynman diagrams for the subprocess of double Higgs boson production via gluon fusion $gg\to hh$.}
\label{diagram:gghh}
\end{figure}

The double Higgs boson production at LHC is dominated by the gluon fusion process, 
$pp\to ggX \to hhX$.
The double Higgs boson production cross section via gluon fusion 
$pp\to hhX$ at LHC with $\sqrt{s}=14$~TeV 
is predicted in the SM as
\begin{align}
\sigma_{\text{SM}}^{\text{NNLO}} (pp \to ggX \to hhX) = 39.1~\text{fb}\;,
\end{align}
where we use 
the PDF of MSTW 2008lo\cite{PDF} and 
the $k$-factor at the NNLO 
$k = 2.4$\cite{deFlorian:2013jea}.
The relevant diagrams in the MCHMs are shown in 
Fig.~\ref{diagram:gghh}.
Deviations in the top Yukawa coupling constant and the triple Higgs boson 
coupling constant affect the prediction of the cross section 
through Fig.~\ref{diagram:gghh}-(a), (b) and (c).
In addition to these contributions, 
due to the existence of the dimension five interaction $hht\bar{t}$, there is a new 
diagram shown in Fig.~\ref{diagram:gghh}-(d).
Therefore, the cross section of this
process depends on the parameters $\kappa_{t}$, $\kappa_{hhh}$, and 
$c_{hhtt}$.
This process is measured by using the decay process of $hh\to \bar{b}b\gamma\gamma $, 
which is expected to be the cleanest mode in the double Higgs boson production\cite{pphh}.

\begin{figure}[t]
  \begin{center}
    \begin{tabular}{cc}
      \resizebox{80mm}{!}{\includegraphics{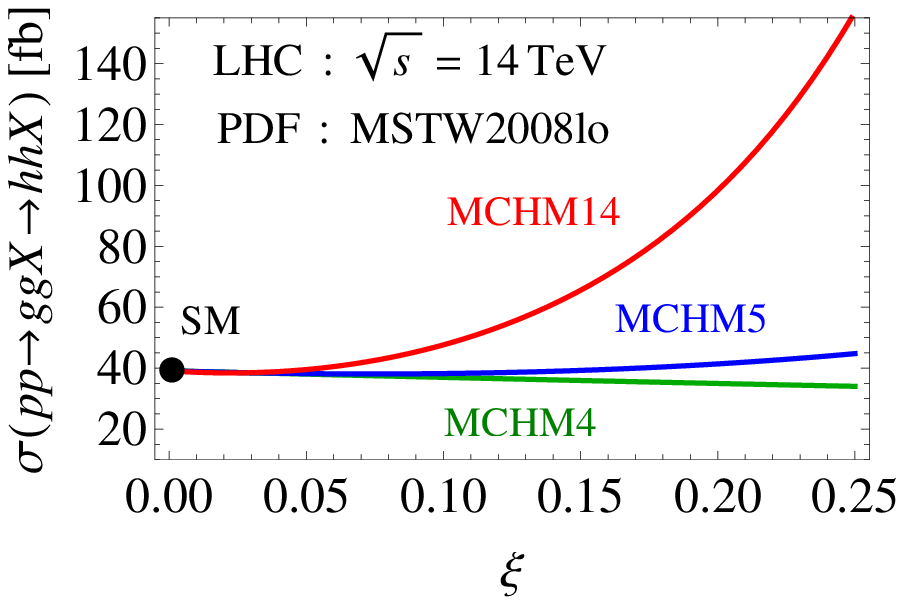}} & 
      \resizebox{80mm}{!}{\includegraphics{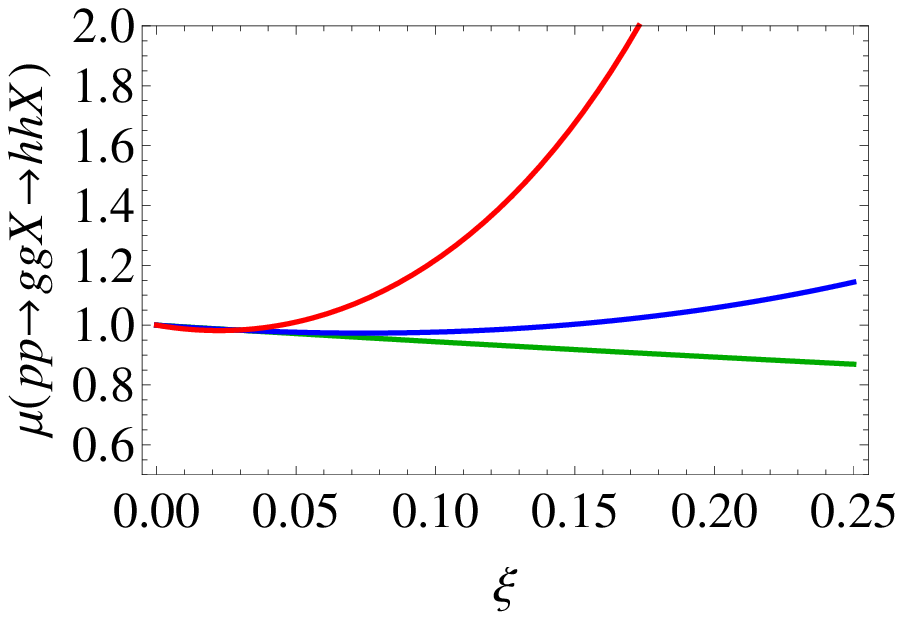}} \\
      (a)&(b)\\[5mm]
      \resizebox{80mm}{!}{\includegraphics{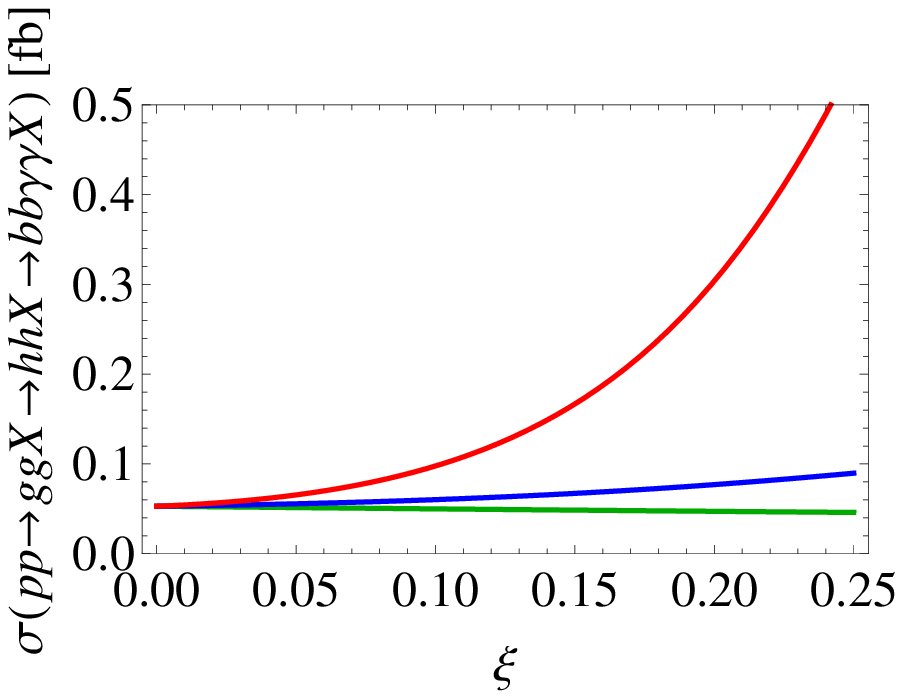}} &
      \resizebox{80mm}{!}{\includegraphics{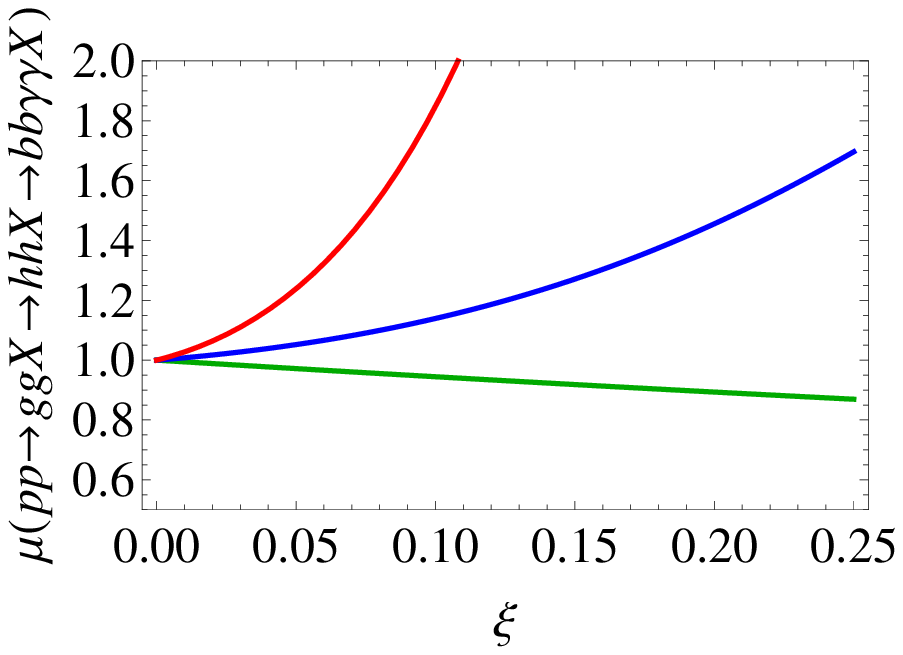}} 
       \\
      (c)&(d)\\
    \end{tabular}
    \caption{(a) The production cross section of $pp \to ggX \to hhX$ in MCHM$4$(green), MCHM$5$(blue) and MCHM${14}$(red)
    at LHC with the collision energy of 14~TeV.
    The $k$-factor of $k = 2.4$ is used.
    (b) The ratio of the production cross section in each model to the SM value.
    (c) The signal cross section of $pp\to ggX\to hhX\to \bar{b}b\gamma\gamma X$ in each model.
    (d) The signal strength of the $pp\to ggX\to hhX\to \bar{b}b\gamma\gamma X$.
    \label{fig:pphh}
    }
  \end{center}
\end{figure}

In Fig.~\ref{fig:pphh}, the production cross section of $pp\to ggX\to hhX$ at the LHC with 
$\sqrt{s}=14$~TeV 
and the ratio of the cross section to the SM prediction 
are shown as a function of the compositeness parameter $\xi$ 
in the three different MCHMs.
We also show the signal cross section $\sigma(pp\to ggX\to hhX\to \bar{b}b\gamma\gamma X)$ and 
the signal strength of this process $\mu$ in each model.
Current upper limit is listed in Table~\ref{constraintxi}, depending on models.
We consider the parameter region of $\xi<0.25$ in MCHM4 and MCHM5,
and $\xi<0.1$ in MCHM14.

In MCHM$4$, the cross section can be 10~\% smaller than the SM prediction 
for $\xi\simeq 0.25$.
In MCHM$5$, an enhancement is found to be 15~\% at $\xi\simeq 0.25$.
In MCHM${14}$, the contribution in Fig.~\ref{diagram:gghh}-(d) 
enhances the production cross section in the large $\xi$ region, because the size of the scale factor 
$c_{hhtt}=-\frac{23}{2}\xi$ is significantly large.
However, in MCHM${14}$ the value of $\xi$ is strongly constrained as 
$\xi< 0.1$, and no significant enhancement is obtained in the allowed region of $\xi$.
The signal cross section in the $\bar{b}b\gamma\gamma$ mode can be enhanced more than 65~\%
in MCHM$5$ for $\xi\simeq 0.25$. 
Even in MCHM${14}$, the signal cross section can be enhanced 
as large as 70~\% in the allowed region ($\xi< 0.1$) due to the significant 
enhancement in $\text{BR}(h\to\gamma\gamma)$.
Although we do not study background by ourselves, which is beyond the scope of our analysis, 
such a large deviation from the SM prediction may be detected 
at the HL-LHC where the SM prediction of the cross section 
of $pp\to hhX\to \gamma\gamma\bar{b}bX$ is expected to be tested 
with the 67\% uncertainty\cite{hhHLLHC}.

\begin{figure}[t]
\begin{center}
\begin{tabular}{cccc}
\includegraphics[scale=0.9]{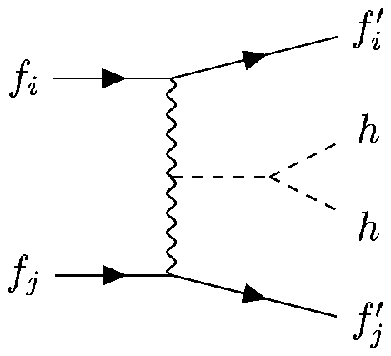}\phantom{Spa}&
\includegraphics[scale=0.9]{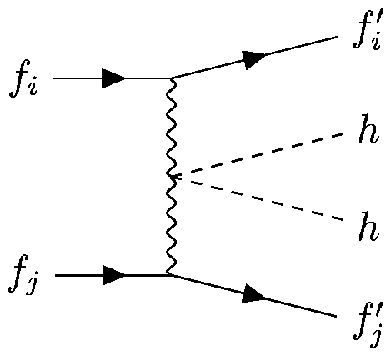}\phantom{Spa}&
\includegraphics[scale=0.9]{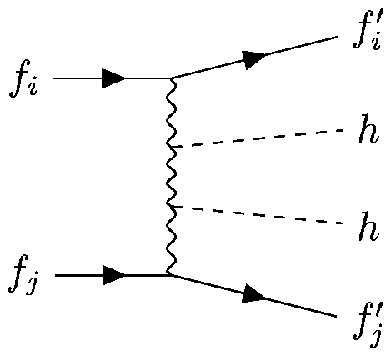}\phantom{Spa}&
\includegraphics[scale=0.9]{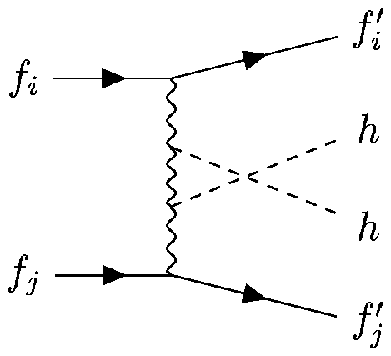}\\
(a)&(b)&(c)&(d)
\end{tabular}
\end{center}
\caption{Relevant diagrams for the double Higgs boson production
process via $W$-fusion. For the external fermions, taking $f_i^{(\prime)},f_j^{(\prime)}=q$ or $\bar{q}$ is for 
the process at LHC and taking $f_i=e^-$, $f_j=e^+$, $f_i'=\nu$ and $f_j'=\bar{\nu}$ is for an electron-positron collider.}
\label{diag:WWtohh}	
\end{figure}

The Higgs boson coupling with weak gauge bosons such as $\kappa_V$ and $c_{hhVV}$ can be
explored by the double Higgs boson production via $W$-fusion.
Relevant diagrams for $W$-fusion at the leading order is shown in Fig.~\ref{diag:WWtohh}.
The elementary process for $W$-fusion is $W^+W^-\to hh$. 
For large centre-of-mass energy of $W^{\pm}$ as $\hat{s}\gg m_W^2$, the amplitude for the various polarisation of $W^{\pm}$ is 
given by 
\begin{equation}
	\mathcal{A}\simeq A_t\frac{\hat{s}}{\hat{t}-m_W^2}+A_u\frac{\hat{s}}{\hat{u}-m_W^2}+A_0+A_s\frac{\hat{s}}{m_W^2}\;,
	\label{eq:AmpWWhh}
\end{equation}
where $A_s$, $A_t$, $A_u$ and $A_0$ are numerical factors given in Table~\ref{WWhhAmp}.
\begin{table}[t]
\caption{
The numerical factors in the scattering amplitude given in Eq.~(\ref{eq:AmpWWhh}).
\label{WWhhAmp}}	
\begin{tabular}{|c|c|c|c|c|}
\hline
Polarisation of $(W^{+},W^{-})$& $A_s$&$A_t$&$A_u$&$A_0$\\	\hline
$(0,0)$&$\frac{g^2}{4}\left(c_{hhVV}-\kappa_V^2\right)$&$\frac{g^2\kappa_V^2}{2}$&$\frac{g^2\kappa_V^2}{2}$&$\frac{g^2}{2}\left(2\kappa_V^2-c_{hhVV}\right)+\frac{g^2m_h^2}{4m_W^2}\left(3\kappa_V\kappa_{hhh}-2\kappa_V^2\right)$
\\ \hline
$(\pm,\pm)$&
0&0&0&$\frac{g^2}{2}\left(c_{hhVV}-\kappa_V^2\right)$\\ \hline
$(\pm,\mp)$&
0&0&0&$-\frac{g^2\kappa_V^2}{2}$\\ \hline
\end{tabular}
\end{table}
In the SM, the cancellation among the diagrams in Fig.~\ref{diag:WWtohh} leads to $A_s=0$ so that
the amplitude is at most constant in the high energy limit.
On the other hand, 
in the MCHMs, the unitarity cancellation does not occur. In such a case, 
perturbative unitarity is expected to be restored at a higher scale $M$, where 
a new heavy resonance with the mass of $M$ may contribute to the $W^+W^-\to hh$ scattering\cite{delayedUnitRe}.
For example, perturbative unitarity is violated at a scale of 2~TeV for $\xi=0.25$ and at a scale of 3~TeV for $\xi=0.1$
\footnote{
At LHC, it is rare for the centre of mass energy of two initial $W$ bosons to be higher than 
2~TeV, due to the PDF suppression. 
Therefore it is not easy to detect the violation of the perturbative unitarity directly even in 
the case of $\xi=0.25$.
}.
As discussed in Ref.~\cite{pphhMCHM,delayedUnitRe}, 
due to this unitarity non-cancellation at the scale $\sqrt{\hat{s}}<M$, 
the scattering cross section of $W^+W^-\to hh$ in the MCHMs are 
enhanced as compared to the SM prediction, even if the relevant coupling constants 
are all suppressed by the scale factors.
Because of this enhancement, 
the production cross section of $pp\to W^+W^-X\to hhX$ is enhanced.

\begin{figure}[t]
  \begin{center}
    \begin{tabular}{ccc}
      \resizebox{80mm}{!}{\includegraphics{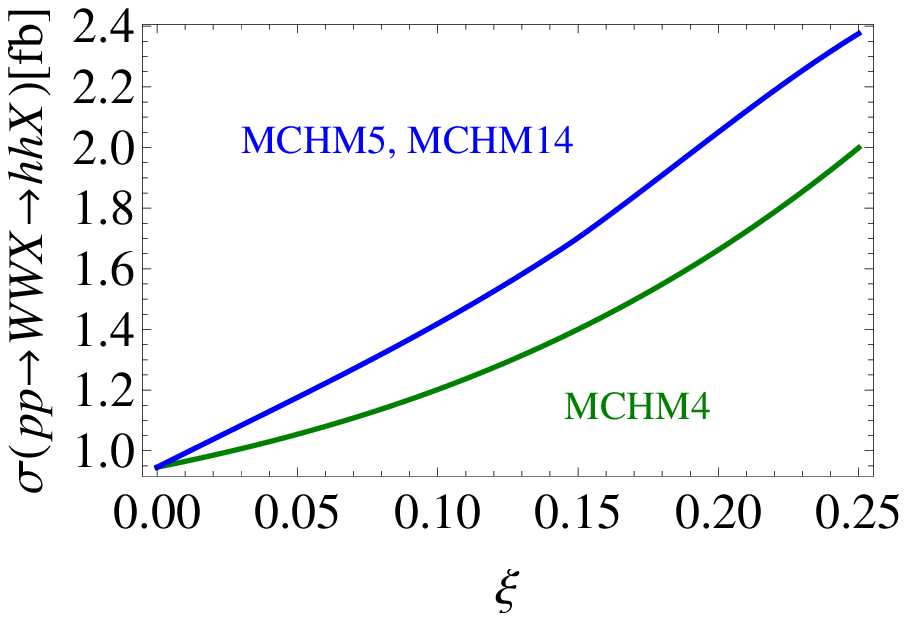}} & 
      \resizebox{80mm}{!}{\includegraphics{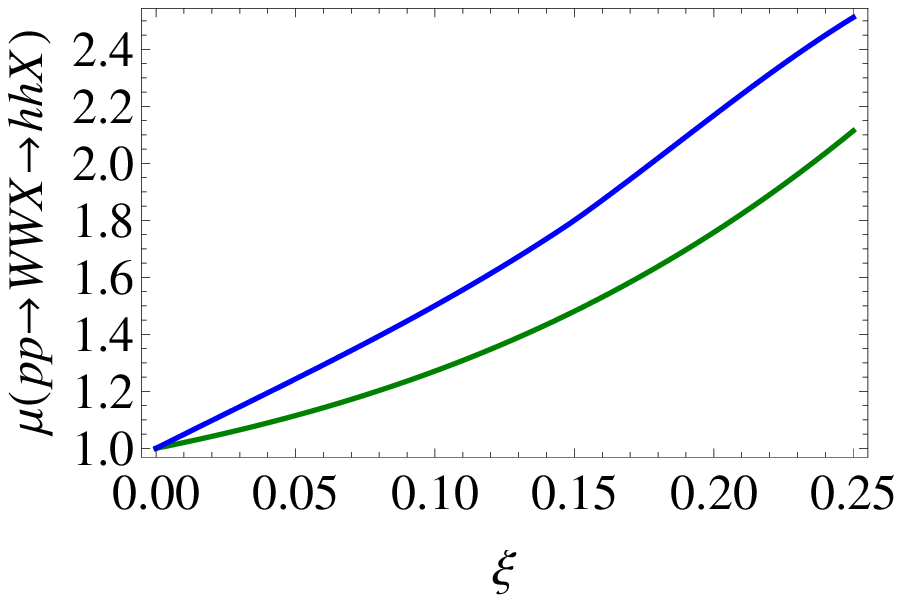}} \\
    \end{tabular}
    \caption{Left: production cross section $pp \to W^+ W^- X \to hh X$ at LHC with the collision energy of 14~TeV as a function of 
    $\xi$. 
    The green curve shows the cross section in MCHM$4$ and the blue curve shows it in 
    the MCHM$5$ and MCHM${14}$. The $k$-factor of $k=1.0$ is used\cite{KfactorWW}.
    Right: The ratio of the cross section in each model to the SM value.}
    \label{WWhhLHC}
  \end{center}
\end{figure}

In Fig.~\ref{WWhhLHC}, we show the production cross section of $pp\to W^+W^-X\to hhX$ in MCHM$4$, MCHM$5$
and MCHM${14}$, 
and the ratio of them to the SM prediction.
Hereafter, we utilise the CalcHep~3.6.25\cite{CalcHep} and the CTEQ6M\cite{CTEQ6M} as the PDF for the numerical computations of the cross section 
of the double Higgs boson production process.
In the evaluation of the cross section, we use the $k$-factor as $k=1.0$\cite{KfactorWW}.
Since the relevant scale factors $\kappa_V$, $c_{hhVV}$, and $\kappa_{hhh}$ 
 in MCHM${14}$ are equal to those in MCHM$5$ as shown in Table~\ref{Table:fp}, 
the production cross section is equal to each other in these two models.
By taking into account the decay modes of the Higgs boson, number of the signal event is
different in MCHM$5$ and MCHM${14}$. 

\begin{figure}[t]
\begin{tabular}{cc}
	\includegraphics[scale=0.8]{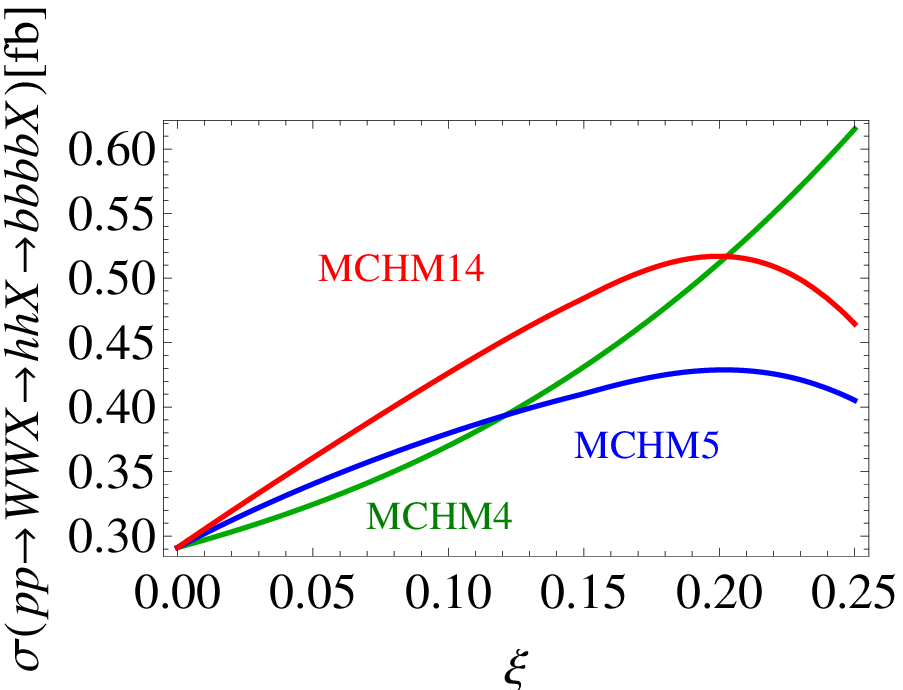}&
	\includegraphics[scale=0.8]{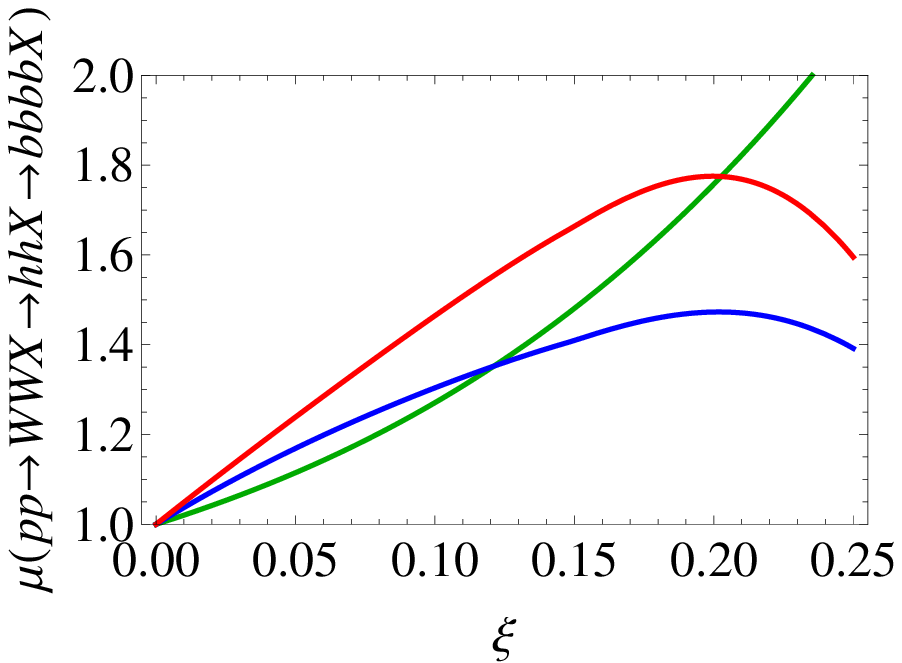}\\
	(a)&(b)\\
	\includegraphics[scale=0.8]{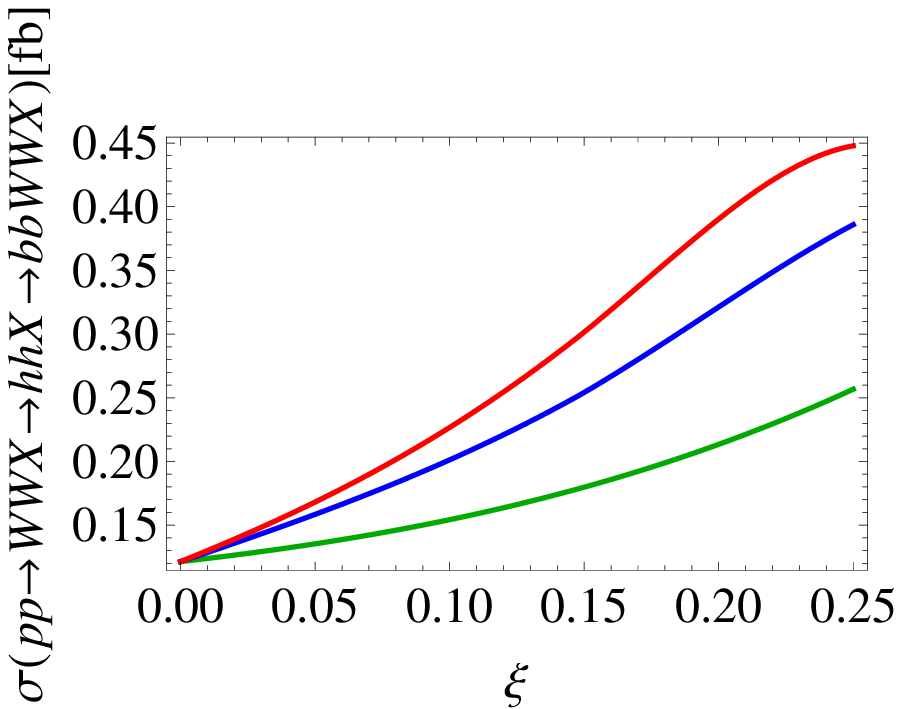}&
	\includegraphics[scale=0.8]{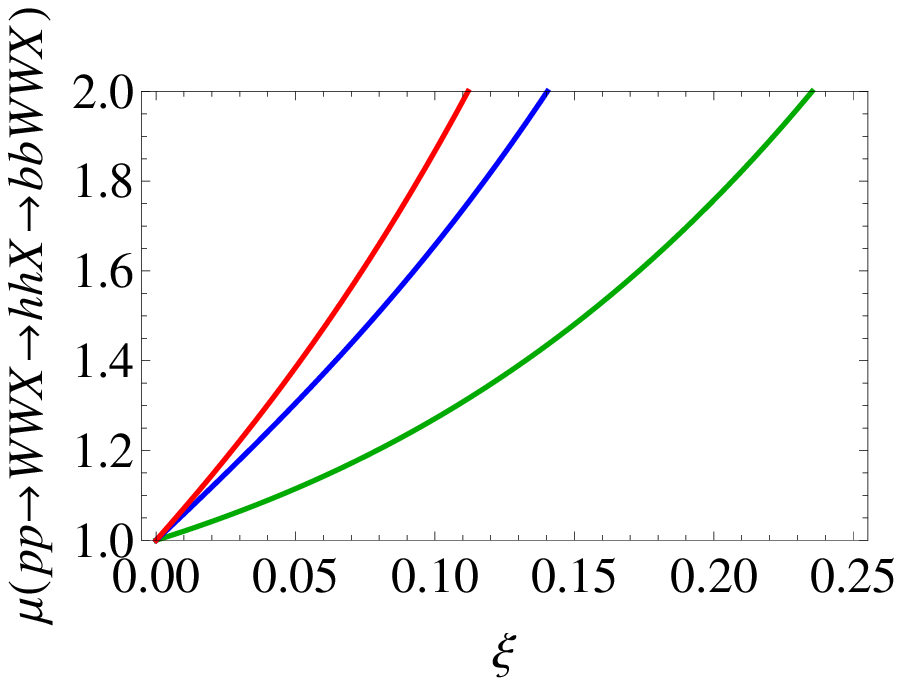}\\
	(c)&(d)\\
\end{tabular}
\caption{
(a) and (b): The signal cross section and the signal strength
at LHC with the collision energy of 14~TeV as 
a function of $\xi$ for 
the double Higgs boson production process via vector boson fusion 
in the decay mode of $hh\to \bar{b}b\bar{b}b$.
(c) and (d): The signal cross section and the signal strength 
as a function of $\xi$ for 
the double Higgs boson production process via vector boson fusion 
in the decay mode of $hh\to \bar{b}bW^+W^-$.
The green, blue and red curves show the predictions in 
MCHM$4$, MCHM$5$ and MCHM${14}$, respectively.
\label{signalstrengthWWhhLHC}
}
\end{figure}

The production cross section of $pp\to W^+W^-X\to hhX$ is as small as a few fb so that  
it is hopeless to detect the process by using the clean decay mode of $hh\to \bar{b}b\gamma\gamma$,
because $\text{BR}(h\to \gamma\gamma)$ is small.
This process is detected in the decay modes of $hh\to \bar{b}b\bar{b}b$ and  $hh\to\bar{b}bW^+W^-$, 
due to rather large branching ratios of $\text{BR}(h\to \bar{b}b)$ and $\text{BR}(h\to W^+W^-)$.
In Fig.~\ref{signalstrengthWWhhLHC}, the signal cross sections 
$\sigma(pp\to W^+W^-X\to hhX\to \bar{b}b\bar{b}bX)$ and $\sigma(pp\to W^+W^-X\to hhX\to \bar{b}bW^+W^-X)$
and their signal strength are shown.
Though the signal cross sections in both modes are of order of 0.1~fb, 
they are larger than the SM predictions.
For example, in MCHM${14}$, the signal cross section for $\xi\simeq 0.1$ is almost twice as large as the SM 
prediction.
Such a large enhancement might be measured at the HL-LHC.

\subsection{Double Higgs boson production at an electron-positron  collider}
The scale factor $\kappa_V$ is very precisely 
measured at future electron-positron linear colliders. For example, the value 
of $\kappa_V$ is determined with accuracy of 1.0~\% by measuring the single Higgs boson
production via $e^+e^-\to Zh$ and $e^+e^-\to h\bar{\nu}\nu$ at the ILC(500) scenario\cite{ILC}. 
In the framework of the MCHMs, it means that the value of the 
compositeness parameter $\xi$ is measured at certain precision.
Not only the gauge coupling of the Higgs boson but also 
the decay branching ratios of the Higgs boson is precisely measured.
For example, the signal strength of the $e^+e^-\to h\bar{\nu}\nu\to \bar{b}b\bar{\nu}\nu$ is 
expected to be measured at 0.7~\% with the collision energy $\sqrt{s}=500$~GeV 
and the integrated luminosity of 500~fb$^{-1}$\cite{ILC}.

The double Higgs boson production is also important to explore the Higgs sector.
This process is sensitive to the triple Higgs boson coupling $hhh$ and the contact interaction 
$hhVV$, which cannot be determined by the single Higgs boson production processes.  
The most promising process for the double Higgs boson production at an electron-positron collider with $\sqrt{s}=500$~GeV is the $Z$ strahlung process as $e^+e^-\to Z^*\to  hhZ$\cite{ZstrahlungatILC,Asakawa:2010xj}.
At the ILC, 
two hopeful decay modes $hh\to \bar{b}b\bar{b}b$ and $hh\to \bar{b}bW^+W^-$ 
are used to detect the double Higgs boson production process. 
We here discuss the production cross section.
The production cross section $e^+e^-\to hhZ$ is measured with accuracy of 42.7~\% (23.7~\%)
at the ILC with $\sqrt{s}=500$~GeV and the integrated luminosity of 500~fb$^{-1}$(1600~fb$^{-1}$)\cite{ILC}.

\begin{figure}[t]
  \begin{center}
    \begin{tabular}{ccc}
      \resizebox{80mm}{!}{\includegraphics{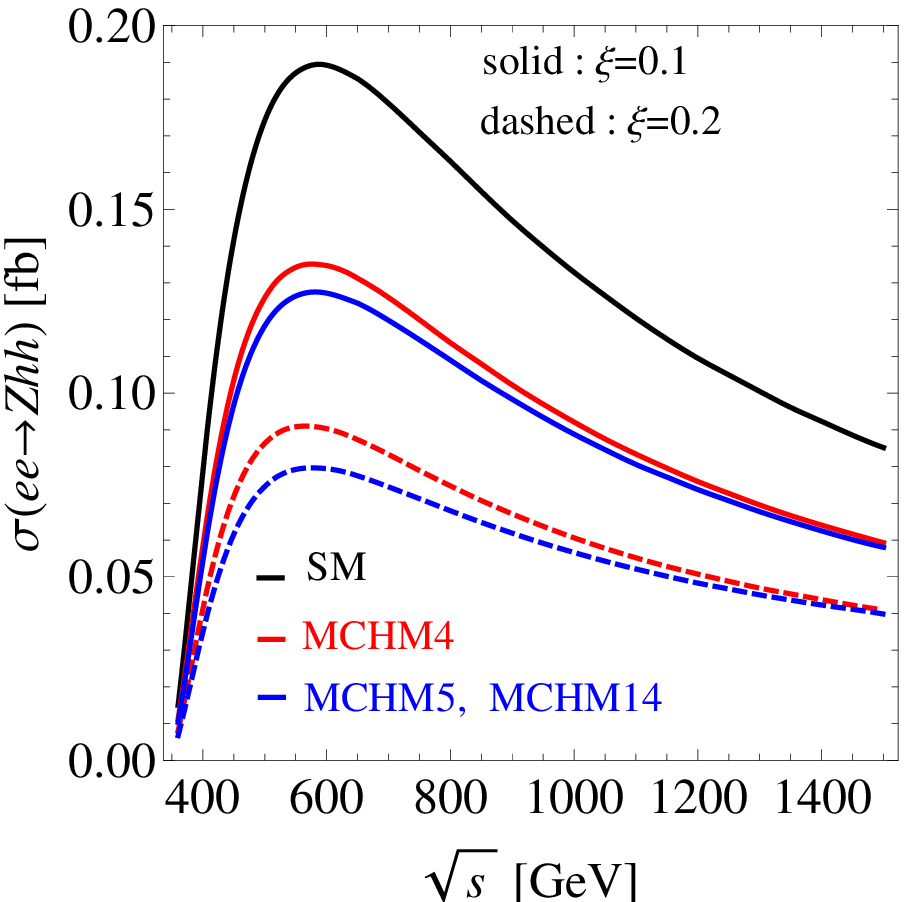}} & 
      \resizebox{80mm}{!}{\includegraphics{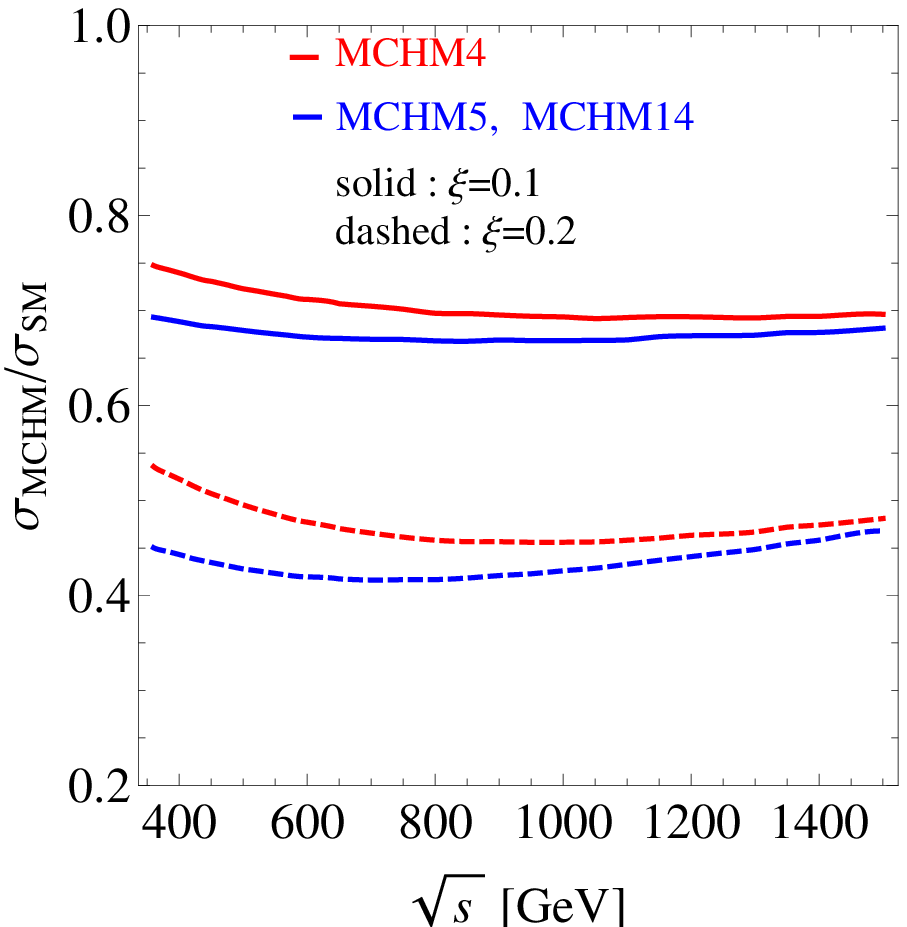}} \\
    \end{tabular}
    \caption{Left: The cross section of $e e \to Z hh$.
The black solid line shows the SM.
The red (blue) solid line shows MCHM$4$ (MCHM$5$ and MCHM${14}$).
The solid (dashed) curves show the case of $\xi = 0.1$ ($\xi = 0.2$).
Right: Ratios of the cross sections to the SM prediction in each model.}
    \label{eeZhh}
  \end{center}
\end{figure}

In Fig.~\ref{eeZhh}, the cross section of $e^+e^-\to Zhh$ and its ratio to  the SM prediction 
is shown in MCHM$4$, MCHM$5$ and MCHM${14}$ as a function of $\sqrt{s}$.
The production cross section in MCHM${14}$ is same as that in MCHM$5$.
In the MCHMs the relevant coupling constants to the decay process of $Z^*\to hhZ$ 
are all suppressed 
by the scale factors $\kappa_V$, $c_{hhVV}$ and $\kappa_{hhh}$  so that the 
production cross section of $e^+e^-\to Z^*\to hhZ$ is suppressed, compared to the SM 
prediction in the all range of $\sqrt{s}$. 
The suppression factor is smaller than 0.5 for $\xi \gtrsim 0.2$.
\begin{figure}[t]
  \begin{center}
    \begin{tabular}{ccc}
      \includegraphics[scale=0.88]{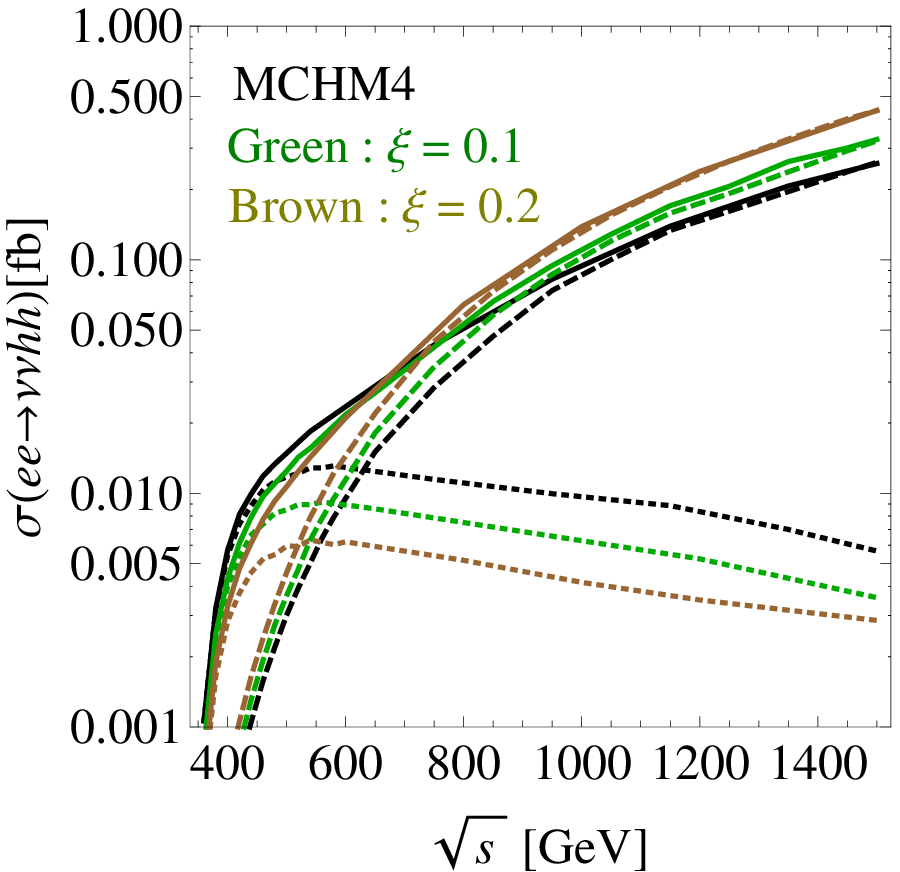}&
  		\includegraphics[scale=0.88]{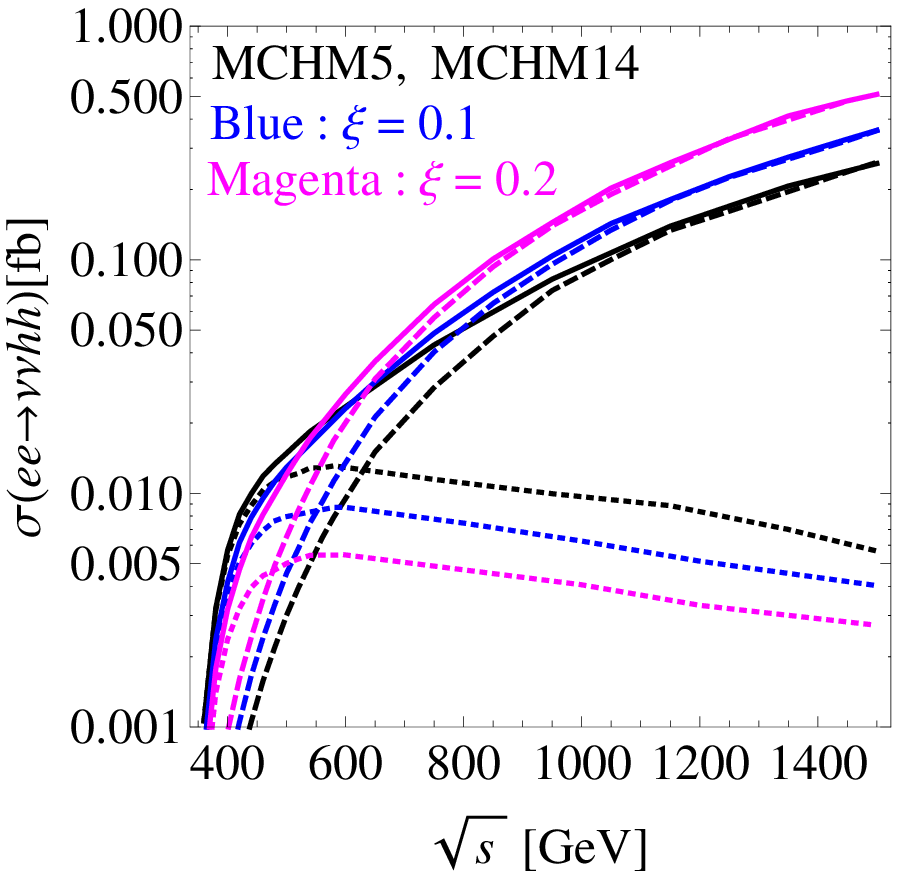}
    \end{tabular}
    \caption{Left: The cross sections for $e^+e^- \to \bar{\nu} \nu hh$ in MCHM$4$. 
    The solid curve is for the total cross section.
   The green and the brown curves are for the case of 
    $\xi=0.1$ and $\xi=0.2$, respectively. 
    Right: The cross section of $e^+e^- \to \bar{\nu} \nu hh$ in
    MCHM$5$ and MCHM${14}$.
    The blue and the magenta curves are for the case of 
    $\xi=0.1$ and $\xi=0.2$, respectively. 
    In the both figures, 
    the dashed and dotted curves show the $W$-fusion and the $Z$-strahlung 
    subprocesses, respectively, and
    the black curves show the SM prediction. 
    }
    \label{eehhnunu}
  \end{center}
\end{figure}
\begin{figure}[t]
  \begin{center}                                     
      \includegraphics[scale=0.9]{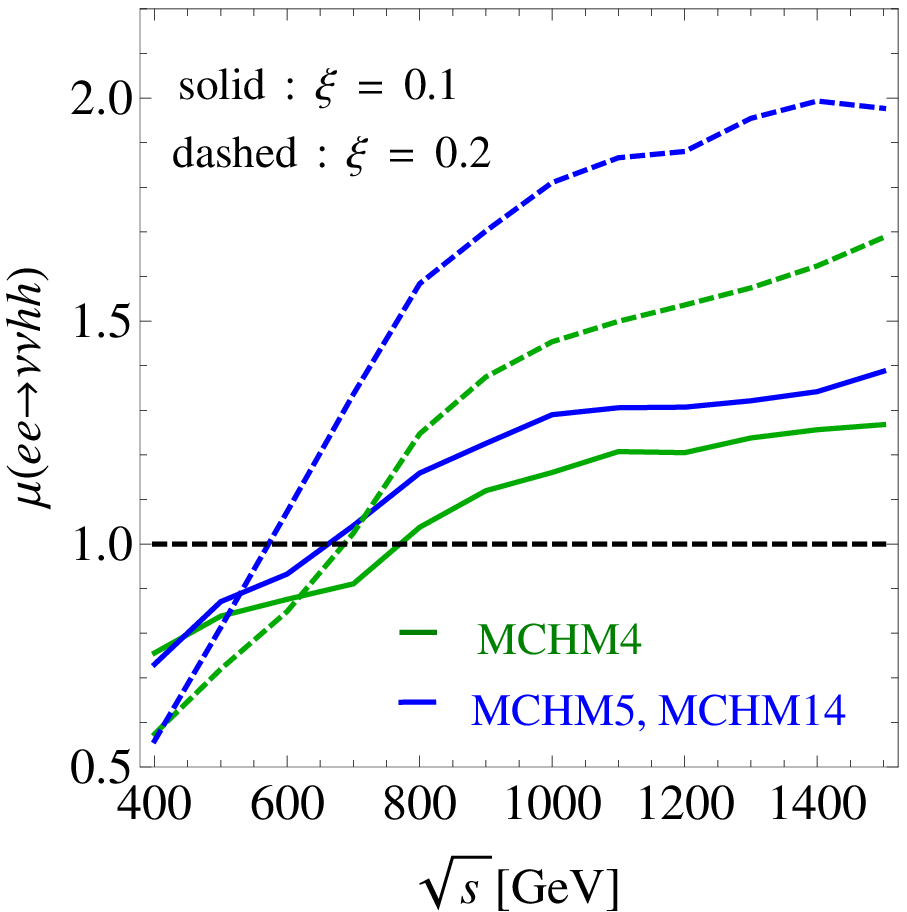}      
    \caption{The ratio of the total cross sections for 
    $e^+e^-\to \bar{\nu}\nu hh$ in the MCHMs to that in the SM.
	The green (blue) curves show the prediction in MCHM$4$ (MCHM$5$ and MCHM${14}$).
    The dashed and dotted curves are for the cases of 
    $\xi=0.1$ and $\xi=0.2$, respectively.
    }
    \label{eehhnunuratio}
  \end{center}
\end{figure}

The double Higgs boson production process with a missing momentum as $e^+e^-\to hh \bar{\nu}\nu$ becomes 
more important for higher $\sqrt{s}$\cite{Asakawa:2010xj,WfusionhhILC}.
It is expected that the production cross section is measured with the accuracy of 26.3~\%(16.7~\%) at 
the ILC with $\sqrt{s}=1$~TeV and the integrated luminosity of 1~ab$^{-1}$ (2.5~ab$^{-1}$)\cite{ILC}. 
This process mainly consists of two different subprocesses; {\it i.e.}, $Z$-strahlung with 
$Z\to \bar{\nu}\nu$ and $W$-fusion as $e^+e^-\to W^+W^-\bar{\nu}\nu\to hh\bar{\nu}\nu$.
The latter provides complementary information to the process of $e^+e^-\to hhZ$.
In Fig.~\ref{eehhnunu}, we show the production cross section of the process $e^+e^-\to hh\bar{\nu}\nu$
in MCHM$4$, MCHM$5$ and MCHM${14}$ for fixed values of the compositeness parameter $\xi=0.1$ and $0.2$ as 
a function of $\sqrt{s}$.
As shown in these figures, the cross section of $e^+e^-\to hh\bar{\nu}\nu$ 
is dominated by $Z$-strahlung for $\sqrt{s}\lesssim 600$~GeV and by $W$-fusion for $\sqrt{s}\gtrsim 600$~GeV. 

In Fig.~\ref{eehhnunuratio}, ratios of the production cross section of $e^+e^-\to hh\bar{\nu}\nu$ to the SM prediction 
are shown 
as a function of $\sqrt{s}$.
As explained before, the $Z$-strahlung process is always suppressed by the scale factors in the MCHMs, on the other 
hand the $W$-fusion process is enhanced in the high energy region because of unitarity non-cancellation.
The cross section of $e^+e^-\to hh\bar{\nu}\nu$ is then suppressed as compared to the 
SM prediction  for $\sqrt{s}\ll 600$~GeV  and it is enhanced for 
$\sqrt{s}\gg 600$~GeV.
Thanks to the expected accuracy of measurements\cite{ILC,CLIC}, 
such a specific behaviour might be observed by the $\sqrt{s}$ scan at the ILC and the CLIC  
unless the compositeness parameter $\xi$ is 
too small.

\begin{figure}[t]
\begin{tabular}{cc}
\includegraphics[scale=0.6]{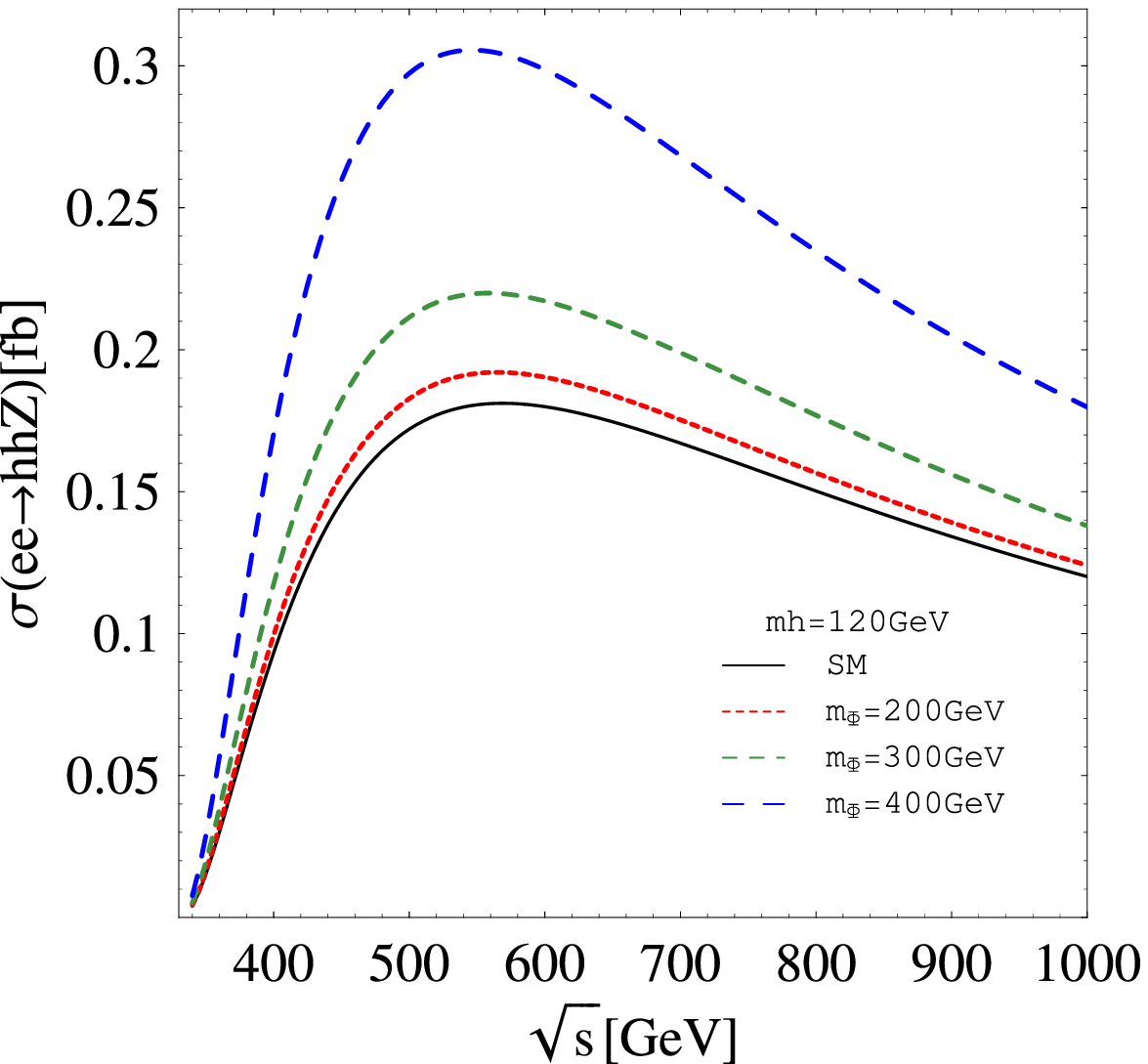}&
\includegraphics[scale=0.6]{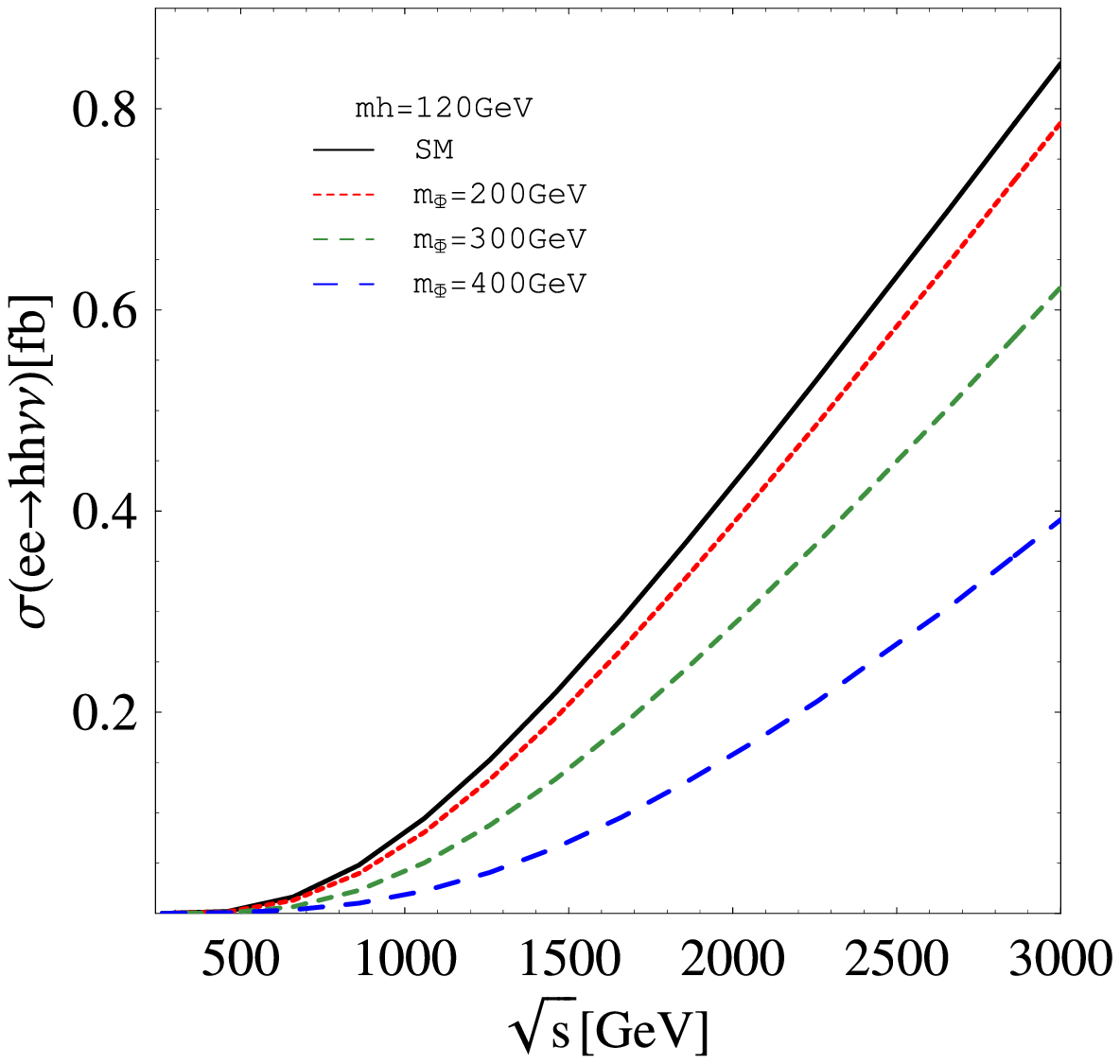}\\
\end{tabular}
	\caption{
	Left: The cross section of $e^+e^-\to hhZ$
	 in the two Higgs doublet model.
	Right: The cross section of $e^+e^-\to hh\bar{\nu}\nu$.
	In the figures, the SM-like Higgs boson mass is fixed as $m_h=120$~GeV and the masses of extra Higgs bosons 
	are taken to be degenerate as $m_{\Phi}\equiv m_{H}=m_A=m_{H^{\pm}}$.
	These figures are taken from Ref.~\cite{Asakawa:2010xj}.
	\label{eehhin2HDM}
	}
\end{figure}

The $\sqrt{s}$ dependence of the double Higgs boson production cross section in the MCHMs 
is different from that in other new physics models.
In Ref.~\cite{Asakawa:2010xj}, the production cross section of $e^+e^-\to hh\bar{\nu}\nu$ 
and $e^+e^-\to hhZ$ are discussed in the new physics models such that 
the triple Higgs boson couplings can significantly deviate from the SM prediction. 
For example, in Fig.~\ref{eehhin2HDM}, which is taken from Ref.~\cite{Asakawa:2010xj}, 
the production cross sections in the two Higgs doublet model
\footnote{
Here we focus on a scenario with the large non-decoupling loop effect\cite{nondecouplingeffect}, 
in which the triple Higgs boson coupling 
is significantly enhanced. 
Such a scenario can play an important
role in a successful electroweak baryogenesis scenario\cite{Kanemura:2004ch}.
}
 are shown.
In the two Higgs doublet model, due to large enhancement of the triple Higgs boson coupling
by loop contributions of the extra Higgs bosons, 
the double Higgs boson production cross section via the $Z$-strahlung process can be enhanced, 
while the double Higgs boson production via $W$-fusion can be suppressed.
In the MCHMs, on the other hand, not only the triple Higgs boson coupling
but also the gauge couplings are deviated from those in the SM.
All the relevant scale factors $\kappa_{hhh}$, $\kappa_V$ and $c_{hhVV}$ are suppressed, 
and the unitarity cancellation is incomplete.
Therefore the $Z$-strahlung process is suppressed, and the $W$-fusion process is enhanced.
This behaviour in the MCHMs is qualitatively different from that in the 
two Higgs doublet model so that these two models can be distinguished from each other.

\section{Conclusion}
We have investigated collider phenomenology of the MCHMs. 
In general, coupling constants of the Higgs boson in the MCHMs can deviate 
from the SM predictions. The deviation pattern depends on the 
compositeness parameter as well as the detail of the matter sector.

We have comprehensively studied how the model can be tested via 
measuring single and double production processes of the Higgs boson
at LHC and future electron-positron colliders.
We have shown the model dependent constraint on the compositeness parameter $\xi$
from the LHC Run-I data. Taking into account the constraint, we have estimated 
the cross section of double Higgs boson production at LHC and electron-positron 
colliders, and 
discussed the possibility to distinguish the matter sector among the MCHMs. 
We have also shown that the $\sqrt{s}$ dependence of the double Higgs boson production cross section
at electron-positron colliders differs from the prediction in other new physics 
models.

In summary, 
at the HL-LHC, a large enhancement of the signal cross section of the double 
Higgs boson production might be measured in some of the MCHMs. 
At the ILC and the CLIC, we can expect that the MCHMs are 
tested with higher precision unless the compositeness parameter 
$\xi$ is too small.
Furthermore, the measurement of the double Higgs boson production process at various $\sqrt{s}$ values would be helpful to discriminate MCHMs from other new physics scenarios.

\begin{acknowledgements}
This work was supported by Grant-in-Aid for Scientific Research from the MEXT, Japan, 
Nos.~23104006 (S. K.) and 23104011 (T. S.), Grant H2020-MSCA-RISE-2014 No.~645722 
(Non Minimal Higgs) (S. K.), and IBS under the project code IBS-R018-D1 (K. K.).
\end{acknowledgements}


\end{document}